\documentclass[aps,prx,reprint,superscriptaddress,floatfix]{revtex4-2}

\pdfoutput=1
\usepackage[T1]{fontenc}
\usepackage{lmodern}
\usepackage[utf8]{inputenc}
\usepackage[english]{babel}

\usepackage{mathtools,bm}
\usepackage{amssymb}
\usepackage{braket}

\usepackage{booktabs,graphicx,microtype}
\graphicspath{{figures/}}

\usepackage{hyperref}

\hypersetup{hidelinks,pdfauthor={Yuting Lyu},pdftitle={BQP-Complete Susceptibilities Observable at Ballistic Times in Local Quantum Dynamics}}
\usepackage{enumitem,etoolbox}
\setlist{leftmargin=*}
\makeatletter
\patchcmd{\@bibdataout@aps}{title="0"}{title="2"}{}{}
\makeatother
\usepackage{amsthm}

\newtheoremstyle{mytheorem}
{6pt}          {6pt}          {\normalfont}  {}             {\itshape}     {.}            {0.5em}        {\thmname{#1}\thmnumber{ #2}\thmnote{ (#3)}}

\theoremstyle{mytheorem}
\newtheorem{theorem}{Theorem}
\newtheorem{corollary}{Corollary}
\newcommand{\C}{\mathbb C}
\newcommand{\R}{\mathbb R}
\newcommand{\T}{\mathbb T}
\newcommand{\Z}{\mathbb Z}
\newcommand{\calH}{\mathcal H}
\newcommand{\calS}{\mathcal S}

\newcommand{\calJ}{\mathcal J}

\newcommand{\BQP}{\mathsf{BQP}}
\newcommand{\Ppoly}{\mathsf{P/poly}}
\DeclareMathOperator{\Tr}{Tr}

\DeclareMathOperator{\supp}{supp}
\DeclareMathOperator{\diag}{diag}
\DeclareMathOperator{\poly}{poly}
\newcommand{\norm}[1]{\left\lVert#1\right\rVert}
\newcommand{\e}{\mathrm e}
\newcommand{\Mfeat}{M_{\rm F}}

\begin{document}

\title{BQP-Complete Susceptibilities Observable at Ballistic Times in Local Quantum Dynamics}

\author{Yuting Lyu}
\email[Contact author: ]{1ytlyu1@gmail.com}
\affiliation{Shenzhen Institute for Quantum Science and Engineering, Southern University of Science and Technology, Shenzhen, China}

\date{\today}

\begin{abstract}
Local response measurements track how perturbations spread in quantum simulators and nonequilibrium many-body systems. Locality bounds constrain the propagation speed, but do not determine whether the response can distinguish computational outcomes at fixed measurement precision as the distance increases. Here we construct time-independent dynamics with bounded nearest-neighbor interactions in which a fixed single-site detector responds to a remote bond phase after a time proportional to their separation. The susceptibility has a positive lower bound independent of distance. With product-state preparations, estimating this susceptibility to constant additive accuracy is complete for bounded-error quantum polynomial time (BQP), with a decision gap independent of distance. Two source-phase settings give the susceptibility with a distance-independent shot count at fixed accuracy and confidence. Inverse-polynomially weak local couplings between the computational register and the chain carrying the excitation preserve the gap. The leading correction vanishes when exchanging the two internal states of the excitation reverses the interaction. Varying the source and observation time makes all bond phases locally identifiable near the zero-phase reference setting. We also prove a conditional quantum advantage for distribution-free response prediction.
\end{abstract}

\maketitle

\section{Introduction}
\label{sec:introduction}
Understanding the propagation of local disturbances is a central problem in nonequilibrium quantum many-body physics and quantum information~\cite{LiebRobinson1972,NachtergaeleSims2006,Cheneau2012LightCone}.
Experiments with ultracold atoms, trapped ions, and superconducting circuits have measured the spread of correlations and quantum information in space and time~\cite{Cheneau2012LightCone,Jurcevic2014Quasiparticle,Braumuller2022InformationPropagation}.
For quantum simulation, a basic question is whether local response functions can be predicted efficiently on a classical computer. 

Baez et~al.~\cite{Baez2020} showed that dynamical correlators and practically accessible variants of dynamical structure factors can retain BQP-hardness. Here BQP denotes the class of decision problems solvable by a quantum computer in polynomial time with bounded error~\cite{BernsteinVazirani1997}.
K\"{o}kc\"{u} et~al.~\cite{Kokcu2024LinearResponse} obtained correlation functions from measured responses to prescribed time-dependent fields.
Piccinelli et~al.~\cite{Piccinelli2026CircuitDifferentiation} expressed retarded Green's functions as circuit derivatives involving inserted perturbations.
We study whether the response to a distant static parameter under fixed, time-independent interactions can distinguish computational outcomes at fixed measurement precision as the distance grows.

Fixed local interactions can execute a quantum computation and produce a readout with a resolvable acceptance gap~\cite{JanzingWocjan2005,NagajWocjan2008,Childs2009}.
Encoding that gap in a remote susceptibility requires the parameter's influence to reach the detector with sufficient amplitude to distinguish accepting from rejecting computations.
Lieb--Robinson bounds constrain the propagation time through an effective light cone~\cite{LiebRobinson1972,NachtergaeleSims2006}.
The signal amplitude within that cone depends on the dynamics. If the susceptibility gap shrinks with distance, resolving the computation requires increasingly precise measurements. A distance-independent gap keeps the required measurement accuracy fixed.

Here we construct time-independent local dynamics in which a susceptibility measured at a fixed single-site detector encodes an autonomous quantum computation with a distance-independent decision gap. For a transport path with $d$ tunable bond phases, a bond at physical distance $\Theta(d)$ from the detector produces a susceptibility with a positive, distance-independent lower bound at an efficiently computable, instance-independent time $t_\star=\Theta(d)$. Here $\Theta(d)$ denotes scaling bounded above and below by positive constants times $d$. With product-state preparations, estimating this susceptibility to constant additive accuracy is BQP-complete on the constructed promise family.

The Hamiltonian and susceptibility problem are defined in Secs.~\ref{sec:unified} and~\ref{sec:active}. Gauge covariance and synchronized state transfer determine the response (Sec.~\ref{sec:transport}), whose stability under local couplings is analyzed in Sec.~\ref{sec:stability-main}. Phase identification and response prediction are treated in Sec.~\ref{sec:extensions}, followed by the transport and interaction requirements in Sec.~\ref{sec:discussion}.

\section{Local Hamiltonian}
\label{sec:unified}

An autonomous computational register executes a program encoded in its initial program symbols. An excitation propagates coherently along a chain of qutrits, carrying phase information to the computational output site. A local measurement at this site combines the output with interference between the excitation's two internal states.

\subsection{Geometry and local degrees of freedom}

Consider a width-two ladder with $\ell$ columns, with the bottom and top rows labeled $\mathrm{bot}$ and $\mathrm{top}$, respectively. The first $L$ sites of the bottom row contain computational registers. Each register consists of a five-dimensional program register and a data qubit, giving a local computational dimension of ten. The program symbols specify local gates and their order, while the data qubits store the state on which these gates act. These registers evolve under fixed nearest-neighbor rules that implement a Hamiltonian quantum cellular automaton (HQCA). To run different programs on the same device, we fix its size and output location for each input length $n$ and compile and pad the circuits to contain the same number $M$ of gates.
Following the register-length choice of Ref.~\cite{NagajWocjan2008}, we take $L=24M$, with the same output column $a\le L$ for all instances of that length.

\begin{figure}[t]
\centering
\includegraphics[width=\columnwidth]{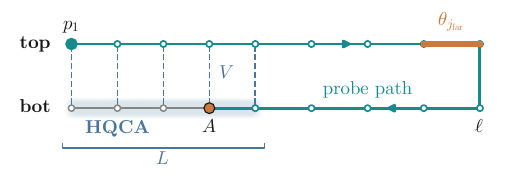}
\caption{Schematic of the width-two ladder and remote readout. The labels top and bot identify the physical rows; the two internal states are the occupied states of each qutrit. The excitation follows the teal route from source $p_1$ to detector $A$, where the computational output and endpoint qutrit are measured jointly. The shaded bottom-row prefix contains the HQCA registers. Blue dashed rungs indicate the local interaction $V$, and $\ell$ labels the rightmost column. The orange bond carries $\theta_{j_{\rm far}}$ at graph distance $\Theta(d)$ from $A$, where $d=m-1$ counts path bonds. Both internal states follow the same route; only hopping in internal state 1 carries Peierls phases.}
\label{fig:unified-ladder}
\end{figure}

The excitation travels along a self-avoiding path $p_1,\ldots,p_m$, as shown in Fig.~\ref{fig:unified-ladder}: it starts at $p_1=(1,\mathrm{top})$, the leftmost site of the top row, runs right to column $\ell$, crosses the rightmost rung, and returns along the bottom row to $A=p_m=(a,\mathrm{bot})$, where $m=2\ell-a+1$. We write $d=m-1$ for the number of path bonds. Extending the path beyond the computational registers adjusts the travel time while keeping the computation and detector location fixed.

The auxiliary chain consists of a qutrit at every path vertex, with a vacuum state $\ket{\mathrm{vac}}$ and two occupied states $\ket0$ and $\ket1$. The spatial row labels $\mathrm{top}$ and $\mathrm{bot}$ are distinct from the excitation's internal-state label $r\in\{0,1\}$. The two internal states propagate along the same physical path and interfere at the endpoint. The detector site contains both the computational output register and a qutrit. The largest local dimension is therefore $10\times3=30$. 

\subsection{Local dynamics}

Each allowed forward transition exchanges a gate symbol with an adjacent blank or pointer.
Exchanging with a blank leaves the data unchanged, whereas exchanging with a pointer applies the corresponding gate to the two data qubits at those sites.
Adding the Hermitian conjugate of each transition gives the time-independent computational Hamiltonian. For neighboring computational columns $u$ and $u+1$, the local update operator is
\begin{equation}
\begin{aligned}
R_{u,u+1}=\sum_{g\in\{W,S,I\}}\big(&
\ket{g,\square}\bra{\square,g}_{\rm prog}\otimes I_4\\
&+\ket{g,\triangleright}\bra{\triangleright,g}_{\rm prog}
\otimes U_g\big)_{u,u+1}.
\end{aligned}
\label{eq:local-update-main}
\end{equation}
Here the program alphabet is $\{\square,\triangleright,W,S,I\}$. $\square$ denotes a blank and $\triangleright$ a pointer. $W,S,I$ are gate symbols of program registers, and the corresponding data gates $U_W,U_S,U_I$ are, respectively, the controlled rotation, swap, and identity gates specified in Appendix~\ref{app:hqca}. The summation label $g$ runs over the three gate symbols. The subscript $\mathrm{prog}$ refers to the two program registers, and $I_4$ is the identity on their two data qubits. For input length $n$, the resulting computational Hamiltonian is
\begin{equation}
\begin{aligned}
H_{{\rm Q},n}&=-\sum_{u=1}^{L-1}\big[\big(R_{u,u+1}+R_{u,u+1}^\dagger\big)\big].
\end{aligned}
\label{eq:computer-main}
\end{equation}
Here $K^{\rm HQCA}_{u,u+1}=-\big(R_{u,u+1}+R_{u,u+1}^\dagger\big)$ ($\|K^{\rm HQCA}_{u,u+1}\|\le2$) is the local computational interaction between the neighboring computational columns $u$ and $u+1$.
For a fixed input length $n$, all computational instances use the same Hamiltonian and differ only in their initial program and data states~\cite{NagajWocjan2008}.

The excitation hops between neighboring path vertices, preserving its internal-state label. For path index $j$ and internal-state label $r\in\{0,1\}$, define the local creation and annihilation operators $a_{j,r}^\dagger=\ket r\bra{\mathrm{vac}}_{p_j}$ and $a_{j,r}=\ket{\mathrm{vac}}\bra{r}_{p_j}$. Internal state $0$ supplies a phase reference, while internal state $1$ acquires the bond phase $\theta_j$ on bond $(p_j,p_{j+1})$. The phase vector is $\bm\theta=(\theta_1,\ldots,\theta_d)\in\T^d$, where $\T=\R/(2\pi\Z)$. The hopping Hamiltonian contains Peierls phase factors~\cite{Peierls1933,Dalibard2011}:
\begin{equation}
\begin{aligned}
H_{\rm P}(\bm\theta)
&=\sum_{j=1}^{d}J_j^{\rm geo}\sum_{r=0}^1
\big(\e^{ir\theta_j}a_{j,r}^\dagger a_{j+1,r}
+\mathrm{h.c.}\big),\\
J_j^{\rm geo}&=\gamma\frac{\sqrt{j(m-j)}}{m}.
\end{aligned}
\label{eq:probe-main}
\end{equation}
Here $J_j^{\rm geo}$ is the hopping amplitude on path bond $j$ in the physical ladder geometry, with a common positive scale factor $\gamma$. The chosen hopping amplitudes permit perfect state transfer (PST) between the path endpoints in the single-excitation sector of $H_{\rm P}(\bm0)$~\cite{Christandl2004,Albanese2004}. 
Each hopping term acts on two neighboring physical sites and conserves the total excitation number in the auxiliary chain. In the single-excitation sector, $\ket{j,r}$ denotes an excitation in internal state $r$ at $p_j$, with every other qutrit in the vacuum. The operator $a_{j,r}^\dagger a_{j+1,r}$ then becomes $\ket{j,r}\bra{j+1,r}$.

Vertical rung couplings connect the auxiliary chain and computational register. Let $W_{u,r}=W_{u,r}^\dagger$ ($\|W_{u,r}\|\le g_{\rm int}$) act on the computational register in column $u$ at $(u,\mathrm{bot})$, which gives the vertical-rung interaction
\begin{equation}
\begin{aligned}
V&=\sum_{u=1}^{L}\sum_{r=0}^1n_{u,r}\otimes W_{u,r}.
\end{aligned}
\label{eq:rung-main}
\end{equation}
Here $n_{j,r}=a_{j,r}^\dagger a_{j,r}$ projects onto internal state $r$ at path vertex $p_j$. In $V$, $n_{u,r}$ acts at $(u,\mathrm{top})$. This interaction conserves excitation number in the auxiliary chain and does not mix the internal states. Appendix~\ref{app:stability} gives a nonzero coupling example and derives the coupling threshold $g_\star$ in Eq.~\eqref{eq:V}.

The time-independent Hamiltonian is
\begin{equation}
H_n^{\rm geo}(\bm\theta)=H_{{\rm Q},n}+H_{\rm P}(\bm\theta)+V.
\label{eq:model-main}
\end{equation}
Each term acts on a physical nearest-neighbor bond and has bounded norm. The Hamiltonian preserves the sector $P_1\calH$ with one excitation in the auxiliary chain. Here $\calH$ is the full tensor-product Hilbert space, and $P_1$ projects onto states with one occupied qutrit and all others in the vacuum. All preparations below lie in this sector. At $V=0$, computation and transport evolve independently. We use this uncoupled system to bound the effect of rung interactions.

\section{Remote response and locality bounds}
\label{sec:active}

\subsection{State preparation and local measurement}

An experimental setting specifies a computational instance, a source site, a relative phase between the internal states, and an evolution time. The initial state is
\begin{equation}
\begin{aligned}
\ket{\psi_{s,\varphi}}&=
\frac{\ket{s,0}+\e^{i\varphi}\ket{s,1}}{\sqrt2},\\
\rho_x&=\rho_z^{\rm Q}\otimes
\ket{\psi_{s,\varphi}}\bra{\psi_{s,\varphi}}.
\end{aligned}
\label{eq:source-main}
\end{equation}
Here $x=(z,s,\varphi,t)$ is the input, $z$ is a length-$n$ computational instance, and $\rho_z^{\rm Q}$ is a computational-basis product state encoding the instance in the program and data registers. The source index satisfies $1\le s\le m$, $\varphi$ is the prepared relative phase between the internal states, and $t$ is the elapsed time. All qutrits except the source are in the vacuum. The superposition of the internal states occupies a single site, so the full preparation remains a product state across sites. The state then evolves under Eq.~\eqref{eq:model-main} for time $t$.

At the endpoint, a fixed local measurement combines the computational output with interference between the amplitudes of the arriving internal states:
\begin{equation}
\begin{aligned}
B_a&=\frac{3I+O_a^{\rm Q}}4,\qquad \frac I2\le B_a\le I.
\end{aligned}
\label{eq:readout-B_a}
\end{equation}
Here $O_a^{\rm Q}$ is the fixed computational output observable, with $\|O_a^{\rm Q}\|\le1$, and $I$ is the identity on the computational register. The affine rescaling in $B_a$ preserves the computational decision gap and gives a nonzero measurement weight for either answer. Define the state $\ket{m,+}:=\frac{\ket{m,0}+\ket{m,1}}{\sqrt2}$. The projector $P_{m,+}=\ket{m,+}\bra{m,+}$ measures interference between the internal states and vanishes on the endpoint vacuum. Both factors act at the same physical site $A$:
\begin{equation}
	\begin{aligned}
		O_A=B_a\otimes P_{m,+},
	\end{aligned}
	\label{eq:readout-O_A}
\end{equation}
where $0\le O_A\le I$. The local response is the probability of this measurement effect:
\begin{equation}
\begin{aligned}
c_{\bm\theta}(x)&=F_{z,s,\varphi}(\bm\theta,t)\\
&=\Tr\!\left[\rho_x\e^{itH_n^{\rm geo}(\bm\theta)}
O_A\e^{-itH_n^{\rm geo}(\bm\theta)}\right].
\end{aligned}
\label{eq:response}
\end{equation}
We use $c_{\bm\theta}(x)$ and $F_{z,s,\varphi}(\bm\theta,t)$ interchangeably, and write $F_x(\bm\theta)$ when the input indices are suppressed. Runs in which the excitation has not reached $A$ contribute zero, so the arrival probability is included in the signal without postselection. The prepared superposition of the internal states and fixed endpoint projector define the phase reference~\cite{BartlettRudolphSpekkens2007,MarvianSpekkens2014}.

\subsection{BQP-complete remote susceptibility}
\label{sec:main-result}

To measure sensitivity to a bond phase, we prepare the endpoint source with relative phase $\pi/2$ and define
\begin{equation}
\chi_z^{(j)}(t):=\left.\partial_{\theta_j}
F_{z,1,\pi/2}(\bm\theta,t)\right|_{\bm\theta=\bm0}.
\label{eq:remote-chi}
\end{equation}
Here $\bm\theta=\bm0$ is the reference setting with all bond phases zero. Only $\theta_j$ is varied, while the input state and detector are held fixed. The source phase $\pi/2$ places the interference signal at a phase-sensitive operating point. The task is to resolve the computational decision gap in this derivative when bond $j$ is far from the detector.

The local Hamiltonian terms and the phase settings used for estimation are known and efficiently computable. Allowed times satisfy $0\le t\le T_n$, where $T_n=\poly(n)$. All inputs have classical descriptions of length polynomial in $n$. The source phases and evolution times can be approximated to additive error $2^{-p}$ in $\poly(n,p)$ time, where $p$ is the requested binary precision. A classical estimator receives the Hamiltonian description and the input $x$. A quantum estimator may simulate the specified dynamics and repeat the fixed local measurement at specified source settings.

\begin{theorem}[BQP-complete remote susceptibility]
\label{thm:unified}
For every BQP-complete promise problem, there is a polynomial-time uniformly constructible family $H_n^{\rm geo}(\bm\theta)$ of polynomial length, with width two, bounded nearest-neighbor terms, and local dimension at most $30$. The Hamiltonian and fixed single-site readout are independent of the promise instance $z$, which is encoded in a product-state preparation. The family includes nonzero, gauge-covariant rung couplings between the computational register and auxiliary chain satisfying $g_{\rm int}\le g_\star$ in Eq.~\eqref{eq:V}.

A bond $j_{\rm far}\in\{1,\ldots,d\}$ lies at physical distance $\Theta(d)$ from the detector and satisfies $\|\partial_{\theta_{j_{\rm far}}}H_n^{\rm geo}\|\ge1/4$. At an efficiently computable observation time $t_\star=\Theta(d)$ independent of $z$, its susceptibility $\chi_z^{\rm far}=\chi_z^{(j_{\rm far})}(t_\star)$ is at least $1/8$. Estimating this susceptibility to additive accuracy $1/100$ is BQP-complete as a promise problem, with decision threshold $3/8$.

For every source setting, the Fourier support of the response is contained in the frequency set of Eq.~\eqref{eq:support}. There are source settings and observation times no later than $t_\star$ for which the response Jacobian obeys Eq.~\eqref{eq:chi-jacobian}. The observation time matches the distance scaling required by locality.
\end{theorem}

Appendix~\ref{app:constants} gives the promise intervals and perturbation allowance. 

\subsection{Lieb--Robinson bound on remote sensitivity}
\label{sec:locality}

Locality constrains when the susceptibility in Eq.~\eqref{eq:remote-chi} can become appreciable. Differentiating the evolution gives an integrated commutator, as in linear-response theory~\cite{Kubo1957}. A Lieb--Robinson estimate~\cite{LiebRobinson1972,NachtergaeleSims2006} then yields
\begin{equation}
\left|\partial_{\theta_j}F_x(\bm\theta,t)\right|
\le B\e^{-\mu(r_j-vt)},\qquad
B=\frac{C_{\rm LR}g_{\rm LR}\|O_A\|}{\mu v}.
\label{eq:lr-gradient}
\end{equation}
Here $r_j=\operatorname{dist}(A,Y_j)$ is the graph distance from the detector to the support $Y_j$ of $G_j=\partial_{\theta_j}H$, and $\|G_j\|\le g_{\rm LR}$. The Hamiltonian $H$ has finite interaction range, bounded graph degree, and a uniformly bounded local interaction sum. $Y_j$ contains a bounded number of sites. The constants $v>0$, $\mu>0$, and $C_{\rm LR}>0$ are the propagation velocity, inverse decay length, and commutator prefactor. They and $B$ are independent of system size. The derivative bound $g_{\rm LR}$ is distinct from the rung coupling strength.

The same estimate constrains simultaneous sensitivity to several phases:
\begin{align}
\sigma_{\min}(\calJ)&\le\sqrt K B\e^{-\mu(r_j-vT)},\label{eq:lr-sv}\\
T&\ge\frac{r_j}{v}-\frac1{\mu v}
\log\!\left(\frac{\sqrt K B}{\kappa_{\rm sens}}\right)
\quad\text{if }\sigma_{\min}(\calJ)\ge\kappa_{\rm sens}.
\label{eq:lr-time}
\end{align}
Here $\calJ_{\nu j}=\partial_{\theta_j}F_{x_\nu}(\bm\theta,t_\nu)$ is the $K\times d$ response Jacobian at a fixed phase point, with $K\ge d$ and $t_\nu\le T$. Its smallest singular value is $\sigma_{\min}(\calJ)$, and $\kappa_{\rm sens}>0$ is a required sensitivity threshold. Equation~\eqref{eq:lr-sv} follows because $\sigma_{\min}(\calJ)$ is bounded above by the Euclidean norm of column $j$, whose entries satisfy Eq.~\eqref{eq:lr-gradient}.
If $r_j=\Theta(d)$, $K\le\poly(d)$, and $\sigma_{\min}(\calJ)\ge\kappa_{\rm sens}\ge1/\poly(d)$, Eq.~\eqref{eq:lr-time} requires $T=\Omega(d)$.
For the same distance scaling, a single susceptibility whose magnitude has a positive lower bound independent of $d$ likewise requires $t=\Omega(d)$.

The propagation bound holds for arbitrary fixed input states, regardless of the complexity of predicting the response. Appendix~\ref{app:universal-lr} gives the proof. Related bounds address correlation generation~\cite{BravyiHastings2006}, perfect state transfer~\cite{Yung2006}, and metrological state preparation~\cite{ChuLiCai2023}.

\section{Gauge covariance and ballistic transport}
\label{sec:transport}

\subsection{Gauge covariance and Fourier support}
\label{sec:gauge}

Only hopping in internal state $1$ depends on the bond phases. These phases can be removed from the hopping terms by a local gauge transformation~\cite{Dalibard2011}:
\begin{equation}
D(\bm\theta)=\prod_{k=1}^{m}\e^{i\Phi_k(\bm\theta)n_{k,1}}.
\label{eq:gauge}
\end{equation}
Here $\Phi_k(\bm\theta)=\sum_{j=k}^d\theta_j$ is the phase accumulated between vertex $k$ and the endpoint, with $\Phi_m=0$, and $n_{k,1}$ is the occupation projector for internal state $1$. Since $\Phi_j-\Phi_{j+1}=\theta_j$, and the computational and rung terms commute with $D$, we have
\begin{equation}
\begin{aligned}
H_n^{\rm geo}(\bm\theta)&=D(\bm\theta)H_n^{\rm geo}(\bm0)D^\dagger(\bm\theta),\\
[O_A,D(\bm\theta)]&=0.
\end{aligned}
\label{eq:cov}
\end{equation}
The detector commutes with the rotation because the endpoint phase is zero. Trace cyclicity therefore transfers all phase dependence to the source:
\begin{equation}
F_{z,s,\varphi}(\bm\theta,t)=F_{z,s,\varphi-\Phi_s(\bm\theta)}(\bm0,t).
\label{eq:shift}
\end{equation}
Here $\Phi_s(\bm\theta)=\bm u_s\cdot\bm\theta=\sum_{j=s}^d\theta_j$, and $(\bm u_s)_j=\mathbf1_{\{j\ge s\}}$ defines the indicator vector $\bm u_s\in\Z^d$ for the bonds from the source to the detector. The indicator $\mathbf1$ equals one when its condition holds and zero otherwise. $\bm u_m=\bm0$ and $\Phi_m=0$.

Expanding the single off-diagonal source coherence gives
\begin{equation}
F_{z,s,\varphi}(\bm\theta,t)=a_{z,s,t}
+2\Re\!\left[\beta_{z,s,t}\e^{i(\varphi-\bm u_s\cdot\bm\theta)}\right].
\label{eq:three}
\end{equation}
Here $a_{z,s,t}\in\R$ is the phase-independent background and $\beta_{z,s,t}\in\C$ is the coefficient of the off-diagonal contribution from the source state. They depend on the instance, source, time, and fixed couplings and are independent of $\varphi$ and $\bm\theta$. Thus every source response has support in the common Fourier frequency set
\begin{equation}
\calS=\{\bm0\}\cup\{\pm\bm u_s:1\le s\le d\},
\qquad |\calS|\le2d+1.
\label{eq:support}
\end{equation}
This support restriction holds at every time and at any strength of the specified gauge-covariant couplings. We use the convention $F(\bm\theta)=\sum_{\bm\ell\in\Z^d}\widehat F(\bm\ell)\e^{i\bm\ell\cdot\bm\theta}$, with normalized phase measure $d^d\bm\theta/(2\pi)^d$. Here $\bm\ell$ is an integer frequency vector and $\widehat F(\bm\ell)$ is its coefficient. At $\varphi=0$ and $s\le d$, the coefficient at $+\bm u_s$ is $\overline{\beta_{z,s,t}}$.

Differentiating Eq.~\eqref{eq:three} gives
\begin{equation}
\chi_z^{(j)}(t)=2\Re\beta_{z,1,t}
=2\Re\widehat F_{z,1,0}(\bm u_1,t),
\label{eq:chi-fourier}
\end{equation}
for every $1\le j\le d$. The same quantity follows from two source-phase settings:
\begin{equation}
\chi_z^{(j)}(t)=\frac12\left[F_{z,1,0}(\bm0,t)-F_{z,1,\pi}(\bm0,t)\right].
\label{eq:two-reference-readout}
\end{equation}
Both settings keep the bond phases at zero and change only the prepared source phase. Gauge covariance relates the bond-phase derivative to a derivative of this source phase, whose single Fourier harmonic gives a two-term parameter-shift rule~\cite{SchuldGradients2019,Wierichs2022}. Repeated bounded measurements estimate the half-difference without numerical differentiation. At fixed additive accuracy and confidence level, the required number of measurement shots is independent of distance by Hoeffding's inequality~\cite{Hoeffding1963}. Uniform quantum simulation gives BQP membership.

More generally, gauge covariance restricts response frequencies to differences between source charge sectors~\cite{MarvianSpekkens2014}:
\begin{equation}
\{\bm q_a-\bm q_b:P_a\rho_xP_b\ne0\}.
\label{eq:general-selection}
\end{equation}
Here $D(\bm\theta)=\sum_a\e^{i\bm q_a\cdot\bm\theta}P_a$, with mutually orthogonal projectors $P_a$ and integer charge vectors $\bm q_a$, and the readout is invariant under $D$. If the source occupies $K$ sectors, the set contains at most $K(K-1)+1$ frequencies. Our source occupies at most two sectors, which restricts the response to at most three frequencies. The coefficients encode the computational dynamics.

\subsection{Computation and interference in the reference system}

To determine the signal amplitude, first set $V=0$. The computational register and auxiliary chain evolve independently, giving
\begin{equation}
\begin{aligned}
F^{(0)}_{z,s,\varphi}(\bm\theta,t)
&=\frac{q_z(t)w_s(t)}2\left[1+\cos(\varphi-\Phi_s(\bm\theta))\right]\\
&=\frac{3+s_z(t)}8\,w_s(t)\left[1+\cos(\varphi-\Phi_s(\bm\theta))\right].
\end{aligned}
\label{eq:unified-response}
\end{equation}
Here $s_z(t)=\Tr[\rho_z^{\rm Q}\e^{itH_{{\rm Q},n}}O_a^{\rm Q}\e^{-itH_{{\rm Q},n}}]$ is the signed computational output, $q_z(t)=(3+s_z(t))/4$ is the expectation of $B_a$, and $w_s(t)=|\bra m\e^{-it\gamma K_m}\ket s|^2$ is the excitation arrival probability. The matrix $K_m$, defined in Eq.~\eqref{eq:main-pst}, describes phase-free, unscaled hopping, and $\ket j$ denotes a path position with the internal-state index suppressed. The superscript $(0)$ denotes the uncoupled reference system.

The decision gap in the computational expectation is multiplied by the arrival probability and the interference fringe. A common observation time with a computational gap and an efficiently computable, constant lower bound on arrival probability therefore yields a constant-gap susceptibility. Inverse-polynomial lower bounds on arrival probabilities for each source site suffice for local phase identification. PST satisfies both transport conditions; the argument also applies to other transport schemes that satisfy them.

\subsection{Synchronization and the computational gap}

The HQCA program motion maps to a free-fermion walk with an instance-independent completion probability~\cite{NagajWocjan2008}. A finite-time spectral average and a polynomial-time search over a rational grid select a common observation time at which
\begin{equation}
s_z(t_\star)\ge\Delta_{\rm sync}\quad(\mathrm{YES}),\qquad
s_z(t_\star)\le-\Delta_{\rm sync}\quad(\mathrm{NO}).
\label{eq:computational-gap-main}
\end{equation}
Here $\Delta_{\rm sync}=97/200$ is the computational output margin. YES and NO label the accepting and rejecting instances of the encoded promise problem, respectively. Circuit padding fixes the HQCA size for each input length, and the selected $t_\star$ is independent of $z$. Appendix~\ref{app:hqca} gives the mapping on the computational subspace, and Appendix~\ref{app:unified} proves the common-time selection.

We match excitation arrival to this time using the hopping matrix of the Krawtchouk chain
\begin{equation}
K_m=\sum_{j=1}^{m-1}J_j(\ket j\bra{j+1}+\mathrm{h.c.}).
\label{eq:main-pst}
\end{equation}
Here $J_j=\sqrt{j(m-j)}/m$ is the unscaled hopping amplitude on path bond $j$. This chain transfers an endpoint excitation perfectly at $t_0=\pi m/2$~\cite{Christandl2004,Albanese2004,Vinet2012}. Extending the ladder beyond the computational prefix and choosing $\gamma=\pi m/(2t_\star)$ makes its physical transfer time equal to $t_\star$. A bond near the far turn lies at physical distance $\Theta(m)$ from the detector and has hopping strength at least $1/4$.

\begin{figure}[t]
\centering
\includegraphics[width=\columnwidth]{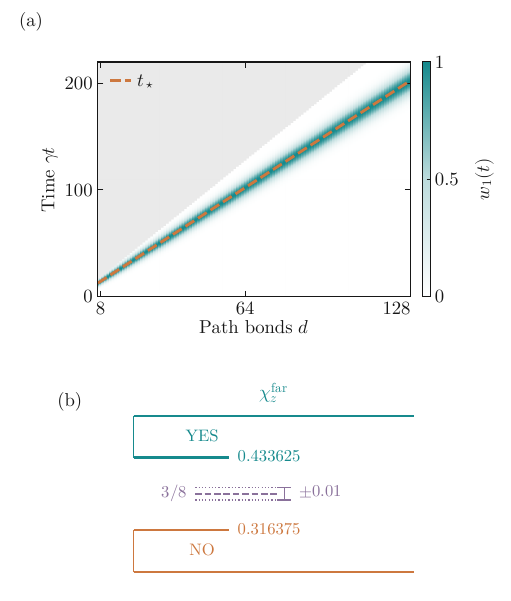}
\caption{Ballistic transfer and a fixed-precision promise gap. (a) Uncoupled arrival probability $w_1(t)=\sin^{2d}[\gamma t/(d+1)]$ for $7\le d\le128$. The orange dashed curve marks first perfect transfer, $\gamma t_\star=\pi(d+1)/2$. Gray shading masks $t>1.25t_\star$. (b) Susceptibility bounds for the allowed coupled family, with $\Delta_{\rm sync}=97/200$ and $\varepsilon_{\rm int}=10^{-3}$. The purple dashed line marks the threshold $3/8$; dotted lines and the bracket show additive tolerance $\pm0.01$. At a time proportional to the remote bond--detector distance, the YES and NO susceptibilities remain distinguishable at fixed additive accuracy.}
\label{fig:ballistic}
\end{figure}

At synchronized arrival, $w_1(t_\star)=1$, as shown in Fig.~\ref{fig:ballistic}(a), so
\begin{equation}
\chi_z^{{\rm far},(0)}=\frac{3+s_z(t_\star)}8,
\qquad
\widehat F^{(0)}_{z,1,0}(\bm u_1,t_\star)=\frac{3+s_z(t_\star)}{16}.
\label{eq:reference-hard-main}
\end{equation}
The output margin in Eq.~\eqref{eq:computational-gap-main} becomes a constant susceptibility gap about $3/8$, as shown in Fig.~\ref{fig:ballistic}(b). 
The geometric rule in Eq.~\eqref{eq:unified-path} keeps the local hopping strengths bounded and gives $t_\star=\Theta(d)$, matching the distance scaling required by locality.
The parameter-shift rule yields a polynomial-time quantum estimator, while estimating the susceptibility to the stated accuracy decides the encoded BQP instance.
The coupling bounds below preserve this margin for nonzero $V$.
The additive-tolerance bracket in Fig.~\ref{fig:ballistic}(b) remains inside the gap between the promise intervals. The displayed YES interval also uses the physical bound $\chi_z^{\rm far}\le1/2$ from Eq.~\eqref{eq:two-reference-readout}.

We compare the engineered and uniform chains at equal path length and maximum bond strength over the same observation window in Fig.~\ref{fig:matched-transport} (Appendix~\ref{app:numeric}).

\section{Stability under local couplings}
\label{sec:stability-main}

\subsection{General rung interactions}

In the single-excitation sector, the rung occupations have orthogonal supports, so $\|P_1VP_1\|\le g_{\rm int}$ regardless of the number of rungs. Duhamel's formula gives
\begin{equation}
|F-F^{(0)}|\le2|t|g_{\rm int},\qquad
|\widehat F-\widehat F^{(0)}|\le2|t|g_{\rm int}.
\label{eq:general-stability-main}
\end{equation}
Here $F$ denotes the response at any fixed input, $\widehat F$ any one of its normalized Fourier coefficients, and $(0)$ the corresponding uncoupled value. Equation~\eqref{eq:chi-fourier} gives $|\chi_z^{\rm far}-\chi_z^{{\rm far},(0)}|\le4t_\star g_{\rm int}$.
For $g_{\rm int}\le g_\star$, the susceptibility error is at most $2\varepsilon_{\rm int}<\Delta_{\rm sync}/8$, so YES and NO susceptibilities remain on opposite sides of the threshold $3/8$. Each source slope changes by at most $1/(16\sqrt m)$, leaving $c_s\ge1/(16\sqrt m)$ and hence
$\sigma_{\min}(\calJ)\ge1/(32\sqrt m)$. The coupling between the occupation number and the data spin in Eq.~\eqref{eq:Vexample} satisfies this restriction and is nonzero.

\subsection{Interactions odd under exchange of the internal states}

When exchanging the two internal states reverses the interaction sign, the first-order correction to the relevant real response vanishes. Consider
\begin{equation}
\begin{aligned}
V_{\rm diff}(g)&=g\sum_{u=1}^L(n_{u,1}-n_{u,0})\otimes X_u,\\
X_u&=X_u^\dagger,\qquad\|X_u\|\le1.
\end{aligned}
\label{eq:quad-V}
\end{equation}
Here $g$ is a signed coupling strength and $X_u$ acts only on the computational register in column $u$. At zero bond phase, the Hamiltonians for the two internal states are
\begin{equation}
\begin{aligned}
\mathcal H_0&=\gamma K_m\otimes I_{\rm Q}+I_m\otimes H_{{\rm Q},n},\\
\mathcal W&=\sum_{u=1}^L\ket u\bra u\otimes X_u,\\
\mathcal H_\pm(g)&=\mathcal H_0\pm g\mathcal W.
\end{aligned}
\label{eq:rail-blocks-main}
\end{equation}
Here $I_m$ and $I_{\rm Q}$ are the identities on the position and computational spaces, and the tensor order is $\C^m\otimes\calH_{{\rm Q},n}$. The signs $+$ and $-$ label internal states $1$ and $0$, respectively. For source $s=1$, define the source-to-detector blocks
\[
A_\pm(g,t)=(\bra m\otimes I_{\rm Q})
\e^{-it\mathcal H_\pm(g)}(\ket1\otimes I_{\rm Q}).
\]
These operators act on the computational space. Their interference term is $\mathcal E(g,t)=\Tr[\rho_z^{\rm Q}A_-^\dagger B_aA_+]$, with the common arguments of $A_\pm$ suppressed. Reversing $g$ exchanges the propagators for the two internal states, giving
\begin{equation}
\mathcal E(-g,t)=\overline{\mathcal E(g,t)},\qquad \partial_g\Re\mathcal E(0,t)=0.
\label{eq:parity-main}
\end{equation}
The Fourier coefficient at $+\bm u_1$ and $\varphi=0$ is $\overline{\mathcal E(g,t)}/4$. Its real part is even in $g$ and its imaginary part can have a linear correction. The full response at $\bm\theta=\bm0$, $\varphi=0$ is also even, since both the preparation and detector are invariant under exchange of the two internal states. Bounds on the second derivatives give the following result, proved in Appendix~\ref{app:quad}.

\begin{corollary}[Quadratic stability under exchange of the internal states]
\label{cor:quadratic}
Let the $X_u$ in Eq.~\eqref{eq:quad-V} be uniformly computable in polynomial time, independent of $z$ and $\bm\theta$, with nonzero choices permitted. If
\begin{equation}
0<|g|\le g_{\rm quad}:=\frac1{t_\star}
\min\!\left\{\sqrt{\frac{\varepsilon_{\rm int}}2},\frac1{4m^{1/4}}\right\},
\label{eq:quad-window}
\end{equation}
the Fourier frequency set is unchanged and
\begin{equation}
\left|\Re\widehat F_{z,1,0}(\bm u_1,t_\star;g)-\frac{3+s_z(t_\star)}{16}\right|
\le\frac{g^2t_\star^2}{2}.
\label{eq:quad-hard-main}
\end{equation}
Here $\varepsilon_{\rm int}=10^{-3}$ is the allowed full-response error. The susceptibility and zero-phase classification response satisfy
\begin{align}
\left|\chi_z^{\rm far}(g)-\frac{3+s_z(t_\star)}8\right|&\le g^2t_\star^2,\label{eq:quad-chi-gap}\\
\left|F_{z,1,0}(\bm0,t_\star;g)-q_z(t_\star)\right|&\le2g^2t_\star^2\le\varepsilon_{\rm int}.
\label{eq:quad-fixed-main}
\end{align}
The BQP-complete susceptibility task, the bound $\chi_z^{\rm far}\ge1/8$, the Jacobian bound, and the learning extension retain their stated thresholds.
\end{corollary}

The sufficient coupling windows for preserving the Jacobian and a fixed response-error budget are of order $1/(\sqrt m\,t_\star)$ for general interactions and $1/(m^{1/4}t_\star)$ for antisymmetric interactions, as shown in Fig.~\ref{fig:stability}(a). Since $t_\star=\Theta(m)$, these windows scale as $m^{-3/2}$ and $m^{-5/4}$. Appendix~\ref{app:universal-leak} bounds Fourier leakage under gauge-breaking perturbations and the resulting prediction error when using the reference features.

\subsection{Finite-size coupled dynamics}
\label{sec:coupled-numerics}

\begin{figure*}[t]
\centering
\includegraphics[width=\textwidth]{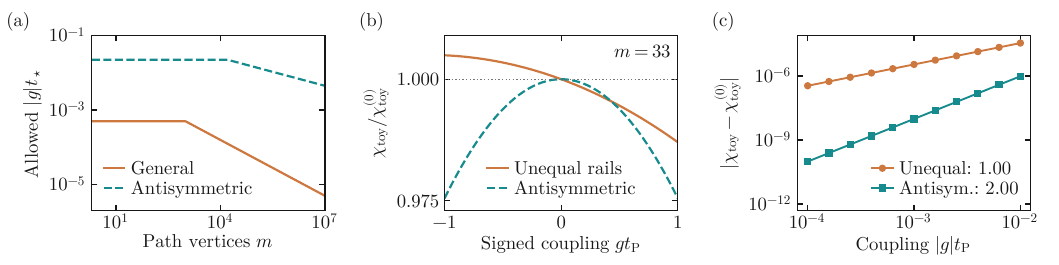}
\caption{Sufficient stability windows and coupled dynamics. (a) Dimensionless upper bounds on the coupling $\lambda=|g|t_\star$ from Eqs.~\eqref{eq:V} and~\eqref{eq:quad-window}, with $\varepsilon_{\rm int}=10^{-3}$. At large $m$, the bounds for general and antisymmetric interactions scale as $m^{-1/2}$ and $m^{-1/4}$, respectively, while maintaining the bounds on the computational gap and the smallest singular value of the response Jacobian. (b) Normalized susceptibility of the three-spin model in Appendix~\ref{app:numeric}, with $m=33$, $\gamma=1.4$, and $t_{\rm P}=\pi m/(2\gamma)$. The dotted horizontal line denotes the uncoupled value one. (c) Numerical absolute errors at the same size and time. Logarithmic fits over $10^{-4}\le|g|t_{\rm P}\le10^{-2}$ give slopes $1.00$ and $2.00$. Exchange symmetry makes the susceptibility even in $g$, eliminating its linear weak-coupling correction.}
\label{fig:stability}
\end{figure*}

For the three-spin register in Appendix~\ref{app:numeric}, we compute the coupled propagators at the state-transfer time $t_{\rm P}$, distinct from the common HQCA completion time $t_\star$. For the antisymmetric interaction, the response is even in the signed coupling [Fig.~\ref{fig:stability}(b)]. At $m=33$, the small-coupling errors have logarithmic slopes $1.00$ and $2.00$ over $10^{-4}\le|g|t_{\rm P}\le10^{-2}$ [Fig.~\ref{fig:stability}(c)]. These calculations show the linear and quadratic response corrections in a finite register. The BQP-completeness proof uses the HQCA construction.

At fixed $g=0.01,0.05$, the responses in Fig.~\ref{fig:fixed-coupling-sizes} sample a finite range of sizes beyond the sufficient stability windows. Size-independent stability and a uniform ordering of the two families' errors require further analysis.

\section{Phase identification and quantum learning}
\label{sec:extensions}

\subsection{Phase sensitivity for each source site}
\label{sec:identifiability}

Hamiltonian identification with restricted access has used measurements at the end of a chain, a small gateway, and measurement time traces~\cite{BurgarthMaruyamaNori2009,BurgarthMaruyama2009,ZhangSarovar2014}. Errors due to unmodeled couplings beyond nearest neighbors have also been analyzed~\cite{GhoshGilchristBurgarth2026}. Other protocols use local observables~\cite{BaireyAradLindner2019} or controlled evolution~\cite{HuangTongFangSu2023}. For a fixed source, the response depends on the sum of bond phases between the source and the detector. Varying the source resolves the individual phase directions locally.

\begin{figure}[t]
\centering
\includegraphics[width=\columnwidth]{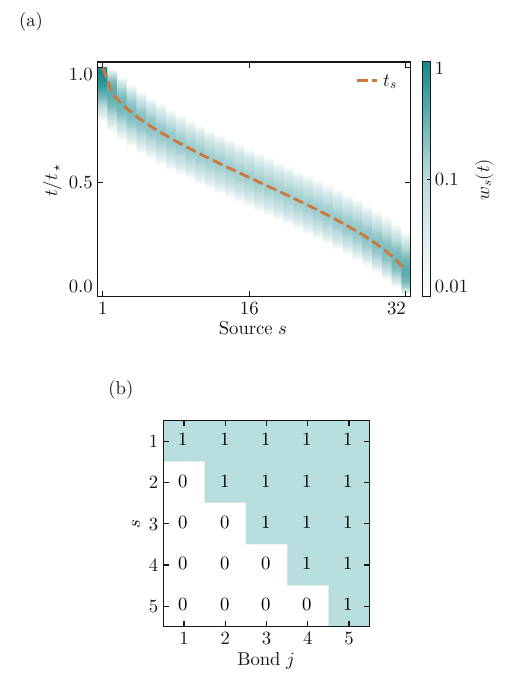}
\caption{Local phase identification using different source sites. (a) Reference arrival probabilities $w_s(t)$ for $m=33$ and $s=1,\ldots,32$. The logarithmic color scale covers $0.01\le w_s\le1$; lower values are white. The orange dashed curve gives $t_s=(m/\gamma)\arcsin\sqrt{(m-s)/(m-1)}\le t_\star$. (b) A $5\times5$ example of $U_{sj}=\mathbf1_{\{j\ge s\}}$. The Jacobian $\mathcal J=\operatorname{diag}(c_s)U$ satisfies $\sigma_{\min}(\mathcal J)\ge[32\sqrt{d+1}]^{-1}$ in the allowed coupling windows. Source-dependent timing thus lets one fixed detector distinguish all bond-phase directions locally near zero phase, assuming a known or calibrated Jacobian.}
\label{fig:identification-learning}
\end{figure}

The time of maximum arrival probability depends on the source [Fig.~\ref{fig:identification-learning}(a)]. At the endpoint transfer time $t_\star$, the uncoupled arrival probability vanishes for every interior source.

At the zero-phase reference point, Eq.~\eqref{eq:three} gives
\begin{equation}
\calJ_{sj}=c_s\mathbf1_{\{j\ge s\}},\qquad c_s=2\Re\beta_{z,s,t_s}.
\label{eq:jacobian-main}
\end{equation}
Here $\calJ$ is the $d\times d$ Jacobian with entries $\calJ_{sj}=\left.\partial_{\theta_j}F_{z,s,\pi/2}(\bm\theta,t_s)\right|_{\bm\theta=\bm0}$, using one source $s=1,\ldots,d$ and a specified observation time $t_s\le t_\star$ per row. The coefficient $c_s$ is the common slope with respect to all bond phases between that source and the detector. The binomial peak bound in Appendix~\ref{app:pst} gives $w_s(t_s)\ge1/(2\sqrt m)$ at the transfer-maximizing times. Since $q_z(t_s)\ge1/2$, the reference slope is at least $1/(8\sqrt m)$, and either coupling window preserves $c_s\ge1/(16\sqrt m)$. Consequently,
\begin{equation}
\calJ=\diag(c_s)U,\qquad
\sigma_{\min}(\calJ)\ge\frac1{32\sqrt m}=\frac1{32\sqrt{d+1}}.
\label{eq:chi-jacobian}
\end{equation}
Here $U_{sj}=\mathbf1_{\{j\ge s\}}$ is the upper-triangular matrix with all entries on and above the diagonal equal to one, as shown in Fig.~\ref{fig:identification-learning}(b). Its inverse takes adjacent differences and satisfies $\|U^{-1}\|_2\le2$. Appendices~\ref{app:pst} and~\ref{app:stability} give the source times and slope bounds. For $s=1$, the same estimates yield $|\partial_{\theta_j}F_{z,1,\pi/2}(\bm0,t_\star)|\ge1/8$ for every bond, including $j_{\rm far}$.

For a known or calibrated Jacobian, Eq.~\eqref{eq:chi-jacobian} bounds the Euclidean noise amplification of linearized inversion by $32\sqrt{d+1}$. One fixed detector therefore distinguishes all phase directions near the reference point using times at most $\Theta(d)$. Extending this local result to global reconstruction requires further analysis, as does determining statistical precision. Fisher information can quantify the latter under suitable regularity assumptions~\cite{KwonLieJiang2026}.

\subsection{Quantum and classical prediction}
\label{sec:learning}

The learning task is to predict the response for a fixed unknown phase vector. Both learners receive the same classical information:
\begin{equation}
\mathcal I_n=\{\operatorname{Desc}(H_n^{\rm geo}(\cdot)),\calS,(X_i,Y_i)_{i=1}^{N_{\rm train}}\}.
\label{eq:shared-information}
\end{equation}
Here $\operatorname{Desc}$ is the known, efficiently computable Hamiltonian description, $\calS$ is the known Fourier frequency set, and $(X_i,Y_i)$ are independent training pairs. Inputs satisfy $X_i\sim\mathcal D$ for an arbitrary distribution on the input domain. Labels have polynomial-length finite-bit encodings, lie in $[0,1]$, and obey $\mathbb E[Y_i\mid X_i=x]=c_{\bm\theta_\star}(x)$, where $\bm\theta_\star$ is the unknown target. Single-bit measurement labels are allowed. For distribution-free, probably approximately correct (PAC) learning~\cite{Valiant1984} and its loss-based regression extension~\cite{Haussler1992}, we use the prediction risk
\begin{equation}
\mathcal R_{\mathcal D}(h)=\mathbb E_{x\sim\mathcal D,\omega}|h(x;\omega)-c_{\bm\theta_\star}(x)|^2.
\label{eq:risk}
\end{equation}
Here $h(x;\omega)$ is the predictor and $\omega$ its fresh evaluation randomness, independent of the test input and training data. The target is the conditional mean response.

Following the Fourier viewpoint on parameter-encoded quantum models~\cite{Schuld2021}, the representation derived in Sec.~\ref{sec:gauge} separates input-dependent features from target-dependent weights:
\begin{equation}
F_x(\bm\theta_\star)=\Re\sum_{\bm\ell\in\calS}b_{\bm\ell}(x)\e^{i\bm\ell\cdot\bm\theta_\star},
\qquad b_{\bm\ell}(x)=\widehat F_x(\bm\ell).
\label{eq:features-main}
\end{equation}
Here the Fourier coefficient $b_{\bm\ell}(x)$ at frequency $\bm\ell$ serves as an input-dependent feature. For $s\le d$, at most three features are nonzero. A quantum learner estimates them by simulating prescribed phase settings $\theta_s=2\pi k/3$, $k=0,1,2$, with all other phases zero, and applying a three-point discrete Fourier transform. For $s=m$, only the constant feature is needed. It then fits the weights by constrained least squares~\cite{BoydVandenberghe2004} and uses fresh quantum feature measurements at prediction time.

For prediction-risk tolerance $\eta\in(0,1)$ and failure probability $\delta\in(0,1)$, the sample count, compilation, training, and per-query prediction resources are polynomial in $n$, $1/\eta$, and $\log(1/\delta)$. The predictor has risk at most $\eta$ with probability at least $1-\delta$ over training. A uniform randomized polynomial-time classical learner with the same distribution-free guarantee and polynomial-size, classically polynomial-time evaluable outputs would imply $\BQP\subseteq\Ppoly$. Here $\Ppoly$ is the class decidable by polynomial-size classical Boolean circuits. The reduction uses the fixed target $\bm\theta_\star=\bm0$ and the common-time computational gap. Appendix~\ref{app:learning-proof} gives the extraction protocol, resource bounds, risk proof, and classical reduction. Appendix~\ref{app:universal-leak} bounds prediction risk with model mismatch for norm-bounded gauge-breaking errors.

Fourier-feature learning and computational separations for dynamics and observables have been studied in Refs.~\cite{Barthe2025,MolteniGyurikDunjko2026}. Here gauge covariance yields a Fourier frequency set of size linear in $d$ while preserving local sensitivity to all $d$ phase directions. Bandyopadhyay et al.~\cite{Bandyopadhyay2026} instead learn a Hamiltonian from short-time samples and simulate it for prediction. Our protocol fits the weights of the response model to scalar labels. Training data can alter computational difficulty~\cite{HuangEtAlNature2021}, and prediction and identification advantages are distinct~\cite{MolteniMarshallDunjko2026}.

\section{Discussion and outlook}
\label{sec:discussion}

The construction encodes a quantum computation in a remote susceptibility with a distance-independent decision gap. It uses a time-independent, width-two ladder with bounded nearest-neighbor interactions, product-state preparations, and a fixed single-site detector. A bond at distance $\Theta(d)$ produces a response derivative with a positive, distance-independent lower bound at an efficiently computable, instance-independent time $t_\star=\Theta(d)$. Estimating this derivative to constant additive accuracy is BQP-complete on the constructed promise family.

Lieb--\allowbreak Robinson bounds set the propagation timescale. Synchronizing perfect state transfer with computation preserves the computational gap at the detector in the reference system. The uniform-XY chain in Appendix~\ref{app:minimal} illustrates the role of transport: its near-detector coefficient remains BQP-complete, while far-source susceptibilities are factorially suppressed in the weak-hopping regime $Jt_\star=1$. The distance-dependent observation time lies beyond the fixed-short-time bounds of cluster-expansion methods for local observables~\cite{WildAlhambra2023}.

Gauge covariance restricts the phase dependence to a known Fourier frequency set of size at most $2d+1$, with the computational decision gap encoded in a single coefficient. The parameter-shift rule extracts the susceptibility directly from local measurement probabilities, including the arrival probability. At fixed accuracy and confidence, the shot count is independent of distance, while each run takes evolution time $t_\star=\Theta(d)$. The prepared coherence between the internal states and endpoint interference provide the reference for measuring the open-path bond phases.

Varying the source and observation time makes all phase directions locally distinguishable at the same detector, with polynomially bounded noise amplification for a known or calibrated Jacobian. Global reconstruction and its measurement cost remain to be determined. The quantum learner estimates the Fourier coefficients and fits the weights using polynomial resources, with fresh quantum measurements at prediction time. Given the same classical training data and model description, a uniform polynomial-time classical learner with the same distribution-free guarantee and polynomial-size outputs that admit polynomial-time classical evaluation would imply $\BQP\subseteq\Ppoly$.

Gauge-covariant couplings preserve this restriction on the Fourier frequencies at any strength. The sufficient coupling bounds for the computational gap and phase sensitivity for each source decrease with system size. The antisymmetric coupling cancels the linear correction to the relevant real response and relaxes the sufficient coupling bound. For norm-bounded gauge-breaking errors, Appendix~\ref{app:universal-leak} bounds Fourier leakage and prediction error using reference features, and susceptibility stability requires additional control of parameter derivatives. The finite-size calculations illustrate the cancellation under exchange of the internal states. Stability at fixed nonzero coupling in the large-system limit remains unresolved.

The observation time is optimal in its scaling with physical distance. Simpler transport profiles and sharper synchronization bounds could reduce the engineering requirements and polynomial overhead. The effects of stronger couplings and decoherence on the susceptibility remain to be determined.

\section*{Data Availability}
The data relevant to this study are publicly available in the Zenodo repository
\cite{lyu_2026_23668910}.

\appendix
\setcounter{secnumdepth}{2}

\section{Autonomous computation and local readout}
\label{app:hqca}

The autonomous computational register uses the ten-state local rules of Ref.~\cite{NagajWocjan2008}. A finite-interval completion bound and a pointwise output relation determine the synchronized observation time.

\subsection{Instance-independent nearest-neighbor Hamiltonian}

Each computational site consists of a five-state program register and a data qubit. The program alphabet is $\mathsf P=\{\square,\triangleright,W,S,I\}$, so the local dimension is ten. The data gates are
\begin{equation}
\begin{gathered}
U_W=\begin{pmatrix}
1&0&0&0\\0&1&0&0\\
0&0&1/\sqrt2&-1/\sqrt2\\
0&0&1/\sqrt2&1/\sqrt2
\end{pmatrix},\\
U_S\ket{x,y}=\ket{y,x},\qquad U_I=I_4.
\end{gathered}
\label{eq:nw-gates}
\end{equation}

Here the matrix for $U_W$ uses the ordered basis $\ket{00},\ket{01},\ket{10},\ket{11}$, and $x,y\in\{0,1\}$ are data bits. The controlled rotation uses $R_y(\pi/2)$, where $R_y(\alpha)=\exp(-i\alpha\sigma_y/2)$, with $\sigma_y$ the Pauli $Y$ matrix. Together with swaps and the standard encoding, these gates support universal computation with polynomial compilation overhead~\cite{NagajWocjan2008}. The local update and Hamiltonian are given in Eqs.~\eqref{eq:local-update-main} and~\eqref{eq:computer-main}. The domain program words of distinct transitions are orthogonal, as are their range words. Since each $U_g$ is unitary, $R_{u,u+1}$ is a norm-one partial isometry, and each computational interaction has norm at most two. We write $N=L$ for the computational chain length, and the family index $n$ denotes input length. The Hamiltonian is fixed for each input length, while the initial product state specifies the computation.

\subsection{Program encoding and computational subspace}

For each input length, compile the BQP instances into nearest-neighbor circuits on $w$ work qubits with $k$ padded rounds. Each round contains one identity delimiter and $w-1$ gates from $\{U_W,U_S,U_I\}$. Thus $M=kw\ge1$ is independent of $z$. Standard bounded-error amplification~\cite{BernsteinVazirani1997} gives
\begin{equation}
 p_z\ge99/100\quad(z\in L_{\rm yes}),\qquad
 p_z\le1/100\quad(z\in L_{\rm no}).
 \label{eq:nw-amplify}
\end{equation}

Here $p_z$ is the acceptance probability, and $L_{\rm yes}$ and $L_{\rm no}$ are the two promise sets. To obtain a sufficient completion probability, embed the $2M$-site HQCA program in $N=24M$ sites. In the original program, the left half contains pointers at $w,2w,\ldots,M$ and blanks elsewhere. The right half contains the $k$ ordered gate rounds. The $w$ work bits occupy original sites $M+1,\ldots,M+w$, while all other data bits start at zero.

Prepend $21M$ blank-program, zero-data cells and append $M$ identity-program, zero-data cells. The $2M$ ordered gate symbols then occupy sites $22M+1,\ldots,24M$, with the first $M$ carrying the compiled circuit. The pointers occupy $21M+w,21M+2w,\ldots,22M$. This specifies a computational-basis product state $\ket{\Phi_z}$. The fixed output is
\begin{equation}
\begin{gathered}
a_N=22M+w,\\
O_{a_N}^{\rm Q}=I_{5,\rm prog}\otimes
 (\ket1\bra1-\ket0\bra0)_{\rm data}.
\end{gathered}
\label{eq:nw-output}
\end{equation}

Here $I_{5,\rm prog}$ is the identity on the program register, and the second factor acts on the output data qubit. In the ladder, $N=L$ and $a=a_N$. The output location and its norm-one observable are fixed for each input length, and the computational preparation is $\rho_z^{\rm Q}=\ket{\Phi_z}\bra{\Phi_z}$.

The update rules preserve the left-to-right order of all $2M$ gate symbols. Their occupied sites $D\subseteq\{1,\ldots,N\}$, with $|D|=2M$, therefore determine the program word. The transition rules of Ref.~\cite{NagajWocjan2008} also fix the data state for each reachable program word: disjoint updates commute, reverse updates undo the corresponding gate, and the pointer and delimiter pattern enforces the circuit order. Denote the resulting basis states by $\ket{\Phi_D}$. Distinct program words make them orthogonal. Each allowed nearest-neighbor gate motion is a $01\leftrightarrow10$ occupation exchange with matrix element $-1$, giving
\begin{equation}
H_{{\rm Q},n}\big|_{\rm legal}\ \simeq\
-\sum_{u=1}^{N-1}
\big(\ket{10}\bra{01}+\ket{01}\bra{10}\big)_{u,u+1}
\bigg|_{|D|=2M}.
\label{eq:nw-hardcore}
\end{equation}

Here $1$ and $0$ indicate presence and absence of a program gate symbol. The restriction to the invariant computational subspace, spanned by these legal configurations, is a hard-core hopping Hamiltonian. Its Jordan--Wigner representation~\cite{JordanWigner1928} consists of $2M$ free fermions initially occupying sites $22M+1,\ldots,N$.

Let $X$ count particles in sites $1,\ldots,22M$. When $X>M$, all first $M$ circuit symbols have crossed the work register, so the output qubit carries the final circuit result. The completion projector $P_{\rm done}$ acts on the program register and commutes with $O_{a_N}^{\rm Q}$. Consequently, completed and unfinished histories contribute without cross terms to the readout.
\subsection{Finite-interval completion bound}

The one-particle hopping matrix for Eq.~\eqref{eq:nw-hardcore} is $h=-\sum_{j=1}^{N-1}(\ket j\bra{j+1}+\mathrm{h.c.})$. Set $K=N+1\ge25$. Its eigenmodes and energies are
\begin{equation}
\begin{aligned}
\phi_r(j)&=\sqrt{\frac2K}\sin\frac{\pi rj}{K},\\
\lambda_r&=-2\cos\frac{\pi r}{K},\quad 1\le r,j\le N.
\end{aligned}
\label{eq:nw-spectrum}
\end{equation}
For any interval $I$ of duration $W$, define $P_I(j\mid k)=W^{-1}\int_I|\langle j|\e^{-ith}|k\rangle|^2dt$. The diagonal spectral contribution is
\begin{equation}
\begin{aligned}
P_\infty(j\mid k)
&=\sum_r|\phi_r(j)|^2|\phi_r(k)|^2\\
&=\frac{1+\frac12\delta_{jk}+\frac12\delta_{j,K-k}}K
\ge\frac1K.
\end{aligned}
\label{eq:nw-pinfty}
\end{equation}
The equality follows from sine-product orthogonality. Consecutive energies obey
\begin{equation}
\begin{aligned}
\lambda_{r+1}-\lambda_r
&=4\sin\frac{(2r+1)\pi}{2K}\sin\frac\pi{2K}\\
&\ge4\sin\frac{3\pi}{2K}\sin\frac\pi{2K}
\ge\frac{12}{K^2},
\end{aligned}
\label{eq:nw-gap}
\end{equation}
using $\sin x\ge2x/\pi$ on $[0,\pi/2]$. Thus every nonzero mode difference has magnitude at least $12/K^2$. Since $|\phi_r(j)|\le\sqrt{2/K}$ and $|W^{-1}\int_I\e^{-it\omega}dt|\le2/(W|\omega|)$,
\begin{equation}
\begin{aligned}
|P_I(j\mid k)-P_\infty(j\mid k)|
&\le\frac2W\sum_{r\ne s}
\frac{|\phi_r(j)\phi_r(k)\phi_s(j)\phi_s(k)|}
{|\lambda_r-\lambda_s|}\\
&\le\frac{2K^2}{3W}.
\end{aligned}
\label{eq:nw-finite}
\end{equation}
For the initial Slater state, the mean occupation $X$ of the leftmost $22M$ sites is a sum of one-particle probabilities. Summing over these sites and the $2M$ initially occupied sites gives
\begin{equation}
\frac1W\int_I\mathbb E_tX\,dt
\ge\frac{44M^2}{K}-\frac{2K^4}{3W}.
\label{eq:nw-X}
\end{equation}
The estimate uses only one-particle transition probabilities.

\subsection{Pointwise computational output}

Completed histories have conditional output $2p_z-1$ and the unfinished contribution has absolute value at most one. Thus
\begin{equation}
\begin{aligned}
s_z(t)&=p_{\rm done}(t)(2p_z-1)\\
&\quad+[1-p_{\rm done}(t)]r_z(t),\qquad |r_z(t)|\le1.
\end{aligned}
\label{eq:nw-decomposition}
\end{equation}
The completion probability depends only on program motion and is independent of the instance. Appendix~\ref{app:unified} selects one common time with $p_{\rm done}\ge3/4$ and applies Eq.~\eqref{eq:nw-decomposition} to obtain the output margin $\Delta_{\rm sync}=97/200$.

\section{Synchronized ladder construction}
\label{app:unified}

The construction in Theorem~\ref{thm:unified} synchronizes computational readout with excitation arrival. The common readout time follows from Appendix~\ref{app:hqca}. The ladder geometry and susceptibility promise intervals follow from the transport and stability bounds in Appendices~\ref{app:pst} and~\ref{app:stability}.

\subsection{Selection of a common observation time}

Clock padding and idling are used in Hamiltonian computation~\cite{CahaLandauNagaj2018}. Synchronization also requires an efficiently computable observation time that is independent of the instance.

For the padded HQCA of Appendix~\ref{app:hqca}, $L=24M$ with $M\ge1$ fixed for each input length. Let $X$ count program particles in the leftmost $22M$ sites and write $p_{\rm done}(t)=\Pr_t(X>M)$, using the completion condition established there. Here $\Pr_t$ and $\mathbb E_t$ denote probability and expectation for the time-$t$ program-position distribution, which is independent of $z$.

Verifier amplification error $\varepsilon_{\rm amp}=1/100$ gives the required output margin whenever $p_{\rm done}(t)\ge p_{\rm min}=3/4$. We first show that a time average exceeds $p_{\rm min}$ by $\delta_{\rm grid}=1/200$, then approximate that average on a rational grid.

The one-particle path eigenvalues are $\lambda_r=2\cos[\pi r/(L+1)]$, with distinct-mode spacing at least $\delta_L=12/(L+1)^2$. On any interval $I$ of duration $W$, each normalized off-diagonal oscillatory integral has magnitude at most $2/(W\delta_L)$. The prefix-occupation calculation in Appendix~\ref{app:hqca}, with $W=(L+1)^6/2$, therefore gives
\begin{align}
\frac1W\int_I\!\mathbb E_t X\,dt
&\ge\frac{44M^2}{L+1}-\frac{4}{3(L+1)^2},\nonumber\\
\frac1W\int_I\!p_{\rm done}(t)\,dt
&\ge\frac{44M}{24M+1}-1\nonumber\\
&\quad-\frac{4}{3M(24M+1)^2}
>p_{\rm min}+\delta_{\rm grid}.
\label{eq:unified-doneavg}
\end{align}

The second inequality uses $0\le X\le2M$ and $\mathbb E_tX\le M[1+p_{\rm done}(t)]$. Its lower bound is smallest at $M=1$, where it equals $0.757866\ldots>p_{\rm min}+\delta_{\rm grid}$.

To select an observation time, take $I=[T/2,T]$ with $T=(L+1)^6$. For the program projector $P_{\rm done}$ onto $X>M$, the bound $\|H_{{\rm Q},n}\|\le2(L-1)$ implies
\begin{equation}
|p_{\rm done}'(t)|\le
\|[H_{{\rm Q},n},P_{\rm done}]\|\le4L.
\end{equation}
Use $Q=\lceil c_{\rm grid}LT\rceil$ equally spaced midpoint times, where $c_{\rm grid}=1000$. The grid-mean error is at most $4L(T/2)/(4Q)\le\delta_{\rm grid}/10$. Hence some grid point has $p_{\rm done}(t)>p_{\rm min}$ with a fixed positive margin. The completion probability at each grid point can be evaluated efficiently from the particle-number distribution of the free fermions.

For the evolved Slater state, Klich's trace formula~\cite{Klich2003} gives the probability-generating function
\begin{equation}
\mathbb E_t[\zeta^X]=\det[I_{2M}-C(t)+\zeta C(t)].
\label{eq:unified-det}
\end{equation}

Here $\zeta$ is a formal variable, $C(t)=B(t)^\dagger B(t)$, and $B(t)$ is the $22M\times2M$ submatrix of $U(t)=\exp(-itA_L)$ connecting initially occupied sites to the left prefix. The matrix $A_L$ is the unit-hopping path adjacency and $U(t)$ is computed from its sine spectrum. Summing the coefficients of $\zeta^{M+1},\ldots,\zeta^{2M}$ gives $p_{\rm done}(t)$. Matrix functions and the determinant polynomial can be evaluated to fixed inverse-polynomial accuracy with polynomially many bit operations. The grid-average surplus ensures that certified numerical intervals identify a qualifying point, including when other points lie near the threshold. The grid size and the evaluation cost are polynomial, so the search runs in uniform polynomial time.

Equation~\eqref{eq:nw-decomposition} bounds the completed contribution using the amplified acceptance probabilities and the unfinished contribution by one. At the selected common time,
\begin{align}
s_z(t_\star)&\ge2(1-\varepsilon_{\rm amp})p_{\rm done}(t_\star)-1\ge\Delta_{\rm sync},&&\text{YES},\\
s_z(t_\star)&\le1-2(1-\varepsilon_{\rm amp})p_{\rm done}(t_\star)\le-\Delta_{\rm sync},&&\text{NO}.
\label{eq:unified-hqca-gap}
\end{align}

The grid search is independent of the instance and therefore gives one observation time for all inputs of the same length.
\subsection{Ladder geometry and transfer time}

Keep the computational register in the first $L$ bottom-row columns, with output at $a\le L$. Extend the ladder and choose
\begin{equation}
\ell=\left\lceil t_\star/3\right\rceil+L,\quad
m=2\ell-a+1,\quad d=m-1,\quad
\gamma=\frac{\pi m}{2t_\star}.
\label{eq:unified-path}
\end{equation}

Here $\ell$ is the number of ladder columns, $m$ the number of path vertices, $d$ the number of phase bonds, and $\gamma$ the common hopping scale. The path vertices run from $p_1=(1,\mathrm{top})$ to $p_\ell=(\ell,\mathrm{top})$, then from $p_{\ell+1}=(\ell,\mathrm{bot})$ back to $p_m=(a,\mathrm{bot})$. Since $t_\star$ is rational, the length rule uses rational comparisons. For $L\ge24$ and $t_\star\ge(L+1)^6/2$, the scale obeys $\pi/3<\gamma<\pi/3+2\pi(L+1)/(L+1)^6<2$ and is efficiently approximable.

Equation~\eqref{eq:probe-main} then has $J_j^{\rm geo}\le\gamma/2<1$, and its reference transfer time is $t_\star$. Choose the top-row bond $j_{\rm far}=\ell-1$. Its phase derivative satisfies
\begin{equation}
\|\partial_{\theta_{j_{\rm far}}}H_{\rm P}\|
=J_{j_{\rm far}}^{\rm geo}\ge\frac14,
\end{equation}
because $j_{\rm far}\ge\ell/2$, $m-j_{\rm far}\ge\ell/2$, and $m\le2\ell$. The bond lies at distance at least $\ell-a-1=\Theta(\ell)=\Theta(m)$ from the detector. At $t_\star$, the excitation in the reference system travels from $p_1$ to $p_m$ with unit probability, and the amplitude in internal state $1$ acquires the factor $\e^{-i\Phi_1}$, where $\Phi_1=\sum_{j=1}^d\theta_j$.

\subsection{Susceptibility gap and error budget}
\label{app:constants}

At synchronized transfer, Eq.~\eqref{eq:unified-response} and the gauge identity~\eqref{eq:chi-fourier} give $\chi_z^{{\rm far},(0)}=[3+s_z(t_\star)]/8$. Set $\varepsilon_{\rm int}=10^{-3}$. The coupling window~\eqref{eq:V} and coefficient bound~\eqref{eq:stability} imply
\begin{equation}
\left|\chi_z^{\rm far}-\frac{3+s_z(t_\star)}8\right|
\le4t_\star g_{\rm int}\le2\varepsilon_{\rm int}.
\label{eq:chi-hard-gap}
\end{equation}
Combining this with Eq.~\eqref{eq:unified-hqca-gap} yields
\begin{equation}
\begin{aligned}
\chi_z^{\rm far}&\ge0.433625\qquad(\mathrm{YES}),\\
\chi_z^{\rm far}&\le0.316375\qquad(\mathrm{NO}).
\end{aligned}
\label{eq:chi-yes-no}
\end{equation}
Each interval is separated from $3/8$ by $0.058625$, so accuracy $1/100$ resolves the encoded BQP decision. Also, $s_z(t_\star)\ge-1$ gives $\chi_z^{\rm far}\ge1/4-2\varepsilon_{\rm int}>1/8$. For BQP membership, Eq.~\eqref{eq:two-reference-readout} estimates the susceptibility using bounded local measurements at source phases $0$ and $\pi$, with polynomial gate cost from the uniform compiler in Appendix~\ref{app:shared-gates}. The susceptibility promise problem is therefore BQP-complete.

For the antisymmetric family, Eq.~\eqref{eq:quad-gradient} replaces the susceptibility error by $g^2t_\star^2\le\varepsilon_{\rm int}/2$ within Eq.~\eqref{eq:quad-window}, preserving the same thresholds. Gauge covariance gives the Fourier-support restriction in Eq.~\eqref{eq:support} for both interaction classes. The arrival-probability and slope estimates for each source in Appendices~\ref{app:pst} and~\ref{app:stability} establish the Jacobian bound at times $t_s\le t_\star$. Since the remote distance and $t_\star$ are both $\Theta(d)$, the construction attains the distance scaling required by the locality bound in Appendix~\ref{app:universal-lr}. All lengths and the allowed nonzero couplings remain polynomially controlled in the input size.

\section{Lieb--Robinson response bound}
\label{app:universal-lr}

A Lieb--Robinson commutator estimate bounds the finite-time phase sensitivity and the response Jacobian, giving a lower bound on the observation time required for remote phase sensitivity.

Consider a differentiable Hamiltonian $K_L(\bm\theta)=\sum_Xh_X(\bm\theta)$ on $L$ sites, where $h_X$ is supported on the finite site set $X$. Assume finite interaction range, bounded graph degree and local dimension, and a uniformly bounded local interaction sum. For $\partial_j=\partial_{\theta_j}$, suppose $G_j=\partial_jK_L$ is supported on a bounded region $Y_j$ and satisfies $\|G_j\|\le g_{\rm LR}$. The Lieb--Robinson estimate~\cite{LiebRobinson1972,NachtergaeleSims2006} is
\begin{equation}
\|[G_j,\tau_u^{\bm\theta}(O)]\|
\le C_{\rm LR}g_{\rm LR}\|O\|
\e^{-\mu(r_j-vu)},\quad u\ge0.
\label{eq:app-lr-assump}
\end{equation}

Here $O$ is supported on a fixed detector region $A$, $r_j=\operatorname{dist}(A,Y_j)$, and $\tau_u^{\bm\theta}(O)=\e^{iuK_L(\bm\theta)}O\e^{-iuK_L(\bm\theta)}$. The constants are uniform in $L$ and $\bm\theta$. Differentiation at fixed input state gives, for $t\ge0$,
\begin{equation}
\partial_j\tau_t^{\bm\theta}(O)
=i\int_0^t\tau_s^{\bm\theta}
\big([G_j,\tau_{t-s}^{\bm\theta}(O)]\big)\,ds.
\label{eq:app-lr-duhamel}
\end{equation}
Unitary invariance of the operator norm, followed by \eqref{eq:app-lr-assump}, yields
\begin{align}
\|\partial_j\tau_t(O)\|
&\le C_{\rm LR}g_{\rm LR}\|O\|\int_0^t
\e^{-\mu(r_j-v(t-s))}\,ds\nonumber\\
&=\frac{C_{\rm LR}g_{\rm LR}\|O\|}{\mu v}
\e^{-\mu r_j}\big(\e^{\mu vt}-1\big)\nonumber\\
&\le B\e^{-\mu(r_j-vt)}.
\label{eq:app-lr-inte}
\end{align}

Here $B=C_{\rm LR}g_{\rm LR}\|O\|/(\mu v)$, with the other constants defined in Sec.~\ref{sec:locality}. Taking an expectation cannot increase the norm. For $K\ge d$ specified settings at times $t_\nu\le T$, define $\calJ_{\nu j}=\partial_j\Tr[\rho_{x_\nu}\tau_{t_\nu}^{\bm\theta}(O)]$. Column $j$ has norm at most $\sqrt K B\exp[-\mu(r_j-vT)]$. Since $\sigma_{\min}(\calJ)\le\|\calJ\bm e_j\|_2$, where $\bm e_j$ is the $j$th coordinate vector, Eq.~\eqref{eq:lr-sv} follows. Solving for $T$ under $\sigma_{\min}\ge\kappa_{\rm sens}$ gives Eq.~\eqref{eq:lr-time}. The result applies to arbitrary fixed input states, including entangled states, under the stated locality hypotheses.

In the ladder, a phase bond near the right-hand turn lies at distance $\Theta(d)$ from $O_A$, with local norms and degree uniformly bounded. For $K=d$ and required sensitivity $\kappa_{\rm sens}=\Omega(d^{-1/2})$, locality therefore requires $T=\Omega(d)$. The synchronized responses attain $t_s\le t_\star=\pi(d+1)/(2\gamma)=\Theta(d)$ with $\sigma_{\min}\ge[32\sqrt{d+1}]^{-1}$, matching this scaling. This bound sets the propagation time required for remote sensitivity.

\section{Spin representation and transfer from each source site}
\label{app:pst}

The spin representation of the engineered hopping chain gives the endpoint transfer probabilities and source-dependent observation times with a uniform inverse-polynomial lower bound on arrival probability.

The hopping matrix of the Krawtchouk chain has a spin-rotation representation~\cite{Christandl2004,Albanese2004}. Identify $\ket j$ with the spin-$J$ basis satisfying $J_z\ket j=(J+1-j)\ket j$, where $J=(m-1)/2$. Then $K_m=(2/m)J_x$, since $\bra jJ_x\ket{j+1}=\tfrac12\sqrt{j(m-j)}$. A $\pi$ rotation exchanges the extreme weights, giving perfect endpoint transfer at $t_0=\pi m/2$. Scaling by $\gamma$ gives $t_\star=t_0/\gamma$.

Representing spin $J$ in the symmetric subspace of $m-1$ spin-$1/2$ particles gives the arrival probability for each source site:
\begin{equation}
 w_s(t)=\binom{m-1}{s-1}
 \sin^{2(m-s)}\!\left(\frac{\gamma t}{m}\right)
 \cos^{2(s-1)}\!\left(\frac{\gamma t}{m}\right).
 \label{eq:binomial-PST}
\end{equation}

For $1\le s\le m-1$, this is maximized at $t_s=(m/\gamma)\arcsin\sqrt{(m-s)/(m-1)}$. The maximum equals the probability of $s-1$ successes in $\operatorname{Bin}(m-1,(s-1)/(m-1))$, where $\operatorname{Bin}(q,p)$ is the binomial distribution with $q$ trials and success probability $p$. The value $s-1$ is a mode. Its probability obeys the uniform bound
\begin{align}
 w_s(t_s)&=\max_{0\le k\le m-1}
 \Pr\{\operatorname{Bin}(m-1,p_s^{\rm bin})=k\}\nonumber\\
 &\ge\frac1{2\sqrt m},\qquad p_s^{\rm bin}=\frac{s-1}{m-1}.
 \label{eq:pst-uniform}
\end{align}
To prove the bound, set $N_{\rm b}=m-1$ and $k=s-1$. For $k=0$, the probability is one. For $1\le k\le N_{\rm b}-1$, applying the Stirling bounds of Robbins~\cite{Robbins1955} for positive integers $q$
\[
\begin{aligned}
 \sqrt{2\pi q}(q/\e)^q&\le q!,\\
 q!&\le\sqrt{2\pi q}(q/\e)^q\e^{1/(12q)},
\end{aligned}
\]
to the binomial probability gives
\begin{align}
 w_s(t_s)&\ge
 \sqrt{\frac{N_{\rm b}}{2\pi k(N_{\rm b}-k)}}
 \e^{-\frac1{12k}-\frac1{12(N_{\rm b}-k)}}\nonumber\\
 &\ge\e^{-1/6}\sqrt{\frac2{\pi N_{\rm b}}}
 >\frac1{2\sqrt m}.
 \label{eq:binomial-peak-bound}
\end{align}
Here $k(N_{\rm b}-k)\le N_{\rm b}^2/4$ and $k,N_{\rm b}-k\ge1$. The case $m=2$ has only $s=1$ and is already covered. For $s=1$, $t_1=t_\star$ and the arrival probability is one. All times and coefficients are efficiently approximable.

Together with Eq.~\eqref{eq:unified-response}, these arrival probabilities give the reference source slope $q_z(t_s)w_s(t_s)/2$ used in the phase-identification bound.

\section{Perturbative response bounds}
\label{app:stability}

Local rung interactions perturb the response probabilities, Fourier coefficients, and phase sensitivities. General couplings give linear error bounds, while antisymmetric interactions yield quadratic bounds for the real parts of Fourier coefficients and selected responses.

\subsection{Interaction norm and general coupling bound}

The rung interaction~\eqref{eq:rung-main} conserves excitation number in the auxiliary chain and commutes with the gauge rotation. In the single-excitation sector, its occupation projectors have orthogonal supports. Consequently,
\begin{equation}
\|P_1VP_1\|=\max_{u,r}\|W_{u,r}\|\le g_{\rm int}.
\label{eq:Vbound}
\end{equation}
Within the single-excitation sector, this bound is independent of the number of rungs. A sufficient coupling window is
\begin{equation}
0<g_{\rm int}\le g_\star:=\min\!\left\{
\frac{\varepsilon_{\rm int}}{2t_\star},
\frac1{64\sqrt m\,t_\star}\right\},
\qquad\varepsilon_{\rm int}=10^{-3}.
\label{eq:V}
\end{equation}
The two terms control the response error and phase sensitivity for each source, respectively. A uniformly computable nonzero example, independent of the instance and phases, is
\begin{equation}
W_{u,0}=-g_\star Z_u^{\rm data},\qquad
W_{u,1}=g_\star Z_u^{\rm data},
\label{eq:Vexample}
\end{equation}
where $Z_u^{\rm data}$ acts as Pauli $Z$ on the data qubit and as the identity on its program register. This interaction couples the occupation number to the data spin across a physical rung and does not commute with hopping in the auxiliary chain. The same example with $g_\star$ replaced by $g$ belongs to the enlarged antisymmetric window~\eqref{eq:quad-window}. The general rung class also permits interactions confined to one internal state, interactions that are symmetric under exchange of the states, and spatially varying interactions.

Set $H^{(0)}=H_{{\rm Q},n}+H_{\rm P}$ and $H=H^{(0)}+V$. Both Hamiltonians and the detector preserve the single-excitation sector containing every preparation. For $U(t)=\e^{-itH}$ and $U_0(t)=\e^{-itH^{(0)}}$, Duhamel's formula gives $\|P_1[U(t)-U_0(t)]P_1\|\le|t|g_{\rm int}$. Since $0\le O_A\le I$,
\begin{equation}
|F-F^{(0)}|\le2|t|g_{\rm int},\qquad
|\widehat F(\bm\ell)-\widehat F^{(0)}(\bm\ell)|\le2|t|g_{\rm int}.
\label{eq:stability}
\end{equation}
The Fourier bound follows by integrating the pointwise estimate against a normalized Fourier character.

\subsection{Exchange symmetry and quadratic stability}
\label{app:quad}

For $V_{\rm diff}(g)$, exchange of the two internal states reverses the sign of $g$. At $\bm\theta=\bm0$, use tensor order $\C^m\otimes\calH_{{\rm Q},n}$ and write $\mathcal H_0=\gamma K_m\otimes I_{\rm Q}+I_m\otimes H_{{\rm Q},n}$. The interaction operator is
\begin{equation}
\mathcal W=\sum_{u=1}^{L}\ket u\bra u\otimes X_u,
\qquad \|\mathcal W\|\le1,
\label{eq:quad-W}
\end{equation}
where the $X_u$ are the Hermitian contractions in Eq.~\eqref{eq:quad-V}. The Hamiltonians for the two internal states are $\mathcal H_\pm(g)=\mathcal H_0\pm g\mathcal W$. For source $s$, their source-to-detector blocks are
\[
A_\pm(g,t)=(\bra m\otimes I_{\rm Q})
\e^{-it\mathcal H_\pm(g)}(\ket s\otimes I_{\rm Q}).
\]
The diagonal contributions from the two internal states are $q_\pm(g,t)$, and their interference term is $\mathcal E(g,t)$:
\begin{equation}
\begin{aligned}
q_\pm(g,t)&=\Tr[\rho_z^{\rm Q}A_\pm^\dagger B_aA_\pm],\\
\mathcal E(g,t)&=\Tr[\rho_z^{\rm Q}A_-^\dagger B_aA_+].
\end{aligned}
\label{eq:quad-interference}
\end{equation}
Gauge covariance and the fixed endpoint projector yield
\begin{equation}
F_{z,s,\varphi}(\bm\theta,t;g)
=\frac{q_-(g,t)+q_+(g,t)}{4}
+\frac12\Re[\e^{i(\varphi-\Phi_s)}\mathcal E(g,t)].
\label{eq:quad-response-exact}
\end{equation}
For $1\le s\le d$, the coefficient of
$\e^{i\bm u_s\cdot\bm\theta}$ at $\varphi=0$ is
$\overline{\mathcal E(g,t)}/4$, whereas the
coefficient $\beta_{z,s,t}$ in Eq.~\eqref{eq:three}
is $\mathcal E(g,t)/4$. Because
$A_\pm(-g,t)=A_\mp(g,t)$ and $B_a$ is Hermitian,
\begin{equation}
\mathcal E(-g,t)=\overline{\mathcal E(g,t)},\qquad
\partial_g\Re\mathcal E(0,t)=0.
\label{eq:quad-parity}
\end{equation}
The two-derivative Duhamel formula gives
$\|\partial_g A_\pm\|\le|t|$ and
$\|\partial_g^2 A_\pm\|\le t^2$. Differentiating the
bilinear expression in Eq.~\eqref{eq:quad-interference} twice and
using $\|B_a\|\le1$ yields
$|\partial_g^2\mathcal E(g,t)|\le4t^2$.
Taylor's theorem, the vanishing first derivative of the real part,
and the reference response~\eqref{eq:unified-response} give
\begin{equation}
\begin{aligned}
|\Re\mathcal E(g,t)-\mathcal E(0,t)|&\le2g^2t^2,\\
\left|\Re\widehat F_{z,s,0}(\bm u_s,t;g)
-\widehat F^{(0)}_{z,s,0}(\bm u_s,t)\right|
&\le\frac{g^2t^2}{2},
\end{aligned}
\label{eq:quad-coeff-stability}
\end{equation}
and in particular
\begin{equation}
\left|2\Re \beta_{z,s,t}(g)-\frac{q_z(t)w_s(t)}2\right|
\le g^2t^2.
\label{eq:quad-gradient}
\end{equation}
Equation~\eqref{eq:quad-coeff-stability} bounds the real part while
the imaginary part may vary linearly with $g$, so the full
complex-coefficient difference can have a linear contribution.

The full response also has a quadratic bound at the zero-phase inputs used in the classical learning reduction.
Let $R$ be the unitary that exchanges the two internal states, $R\ket{j,0}=\ket{j,1}$ and $R\ket{j,1}=\ket{j,0}$, acting trivially on computational registers. At $\bm\theta=\bm0$,
$RH(g)R^\dagger=H(-g)$. Both the $\varphi=0$ source and
$O_A=B_a\otimes P_{m,+}$ are invariant under $R$. Hence
$F_{z,s,0}(\bm0,t;g)$ is even in $g$.
Two Duhamel derivatives of the full-state expectation obey
$|\partial_g^2F_{z,s,0}(\bm0,t;g)|\le4t^2$, since the
perturbation norm in the single-excitation sector is $\le1$ per unit $g$.
Consequently,
\begin{equation}
|F_{z,s,0}(\bm0,t;g)-F^{(0)}_{z,s,0}(\bm0,t)|
\le2g^2t^2.
\label{eq:quad-response}
\end{equation}

The full-response symmetry holds at zero bond phase and $\varphi=0$. At $s=1$ and $t=t_\star$, Eq.~\eqref{eq:quad-response} gives the classification-response bound~\eqref{eq:quad-fixed-main} used in the learning reduction.

\subsection{Phase sensitivity for each source site and finite precision}

For the general rung class, gauge covariance converts coefficient stability into derivative stability. Equation~\eqref{eq:three} holds for every source and time, including different interactions for the two internal states. At $\varphi=\pi/2$ and $\bm\theta=\bm0$,
\begin{equation}
 \partial_{\theta_j}F_{z,s,\pi/2}(\bm0,t)
 =2\Re \beta_{z,s,t}\,\mathbf1_{\{j\ge s\}}.
\label{eq:derivative-stability}
\end{equation}
For $1\le s\le d$, the coefficient of $\exp(i\bm u_s\cdot\bm\theta)$ at
$\varphi=0$ is $\overline{\beta_{z,s,t}}$, so
Eq.~\eqref{eq:stability} and the decoupled response
Eq.~\eqref{eq:unified-response} imply
\begin{equation}
 \left|2\Re \beta_{z,s,t}-\frac{q_z(t)w_s(t)}2\right|
 \le4|t|g_{\rm int}.
\label{eq:coefficient-gradient}
\end{equation}
The bound is independent of the number of phase coordinates.

For each source $s=1,\ldots,d$, choose the maximizing time $t_s\le t_\star$ from Appendix~\ref{app:pst}. The reference slope obeys $c_s^{(0)}=q_z(t_s)w_s(t_s)/2\ge1/(8\sqrt m)$. Equations~\eqref{eq:coefficient-gradient} and~\eqref{eq:V} give $|c_s-c_s^{(0)}|\le1/(16\sqrt m)$. For the antisymmetric family, Eqs.~\eqref{eq:quad-gradient} and~\eqref{eq:quad-window} instead give
\[
|c_s-c_s^{(0)}|\le g^2t_s^2\le g^2t_\star^2\le\frac1{16\sqrt m}.
\]
Thus both classes retain $c_s\ge1/(16\sqrt m)$. The Jacobian is
\begin{equation}
\calJ=\diag(c_s)U,\qquad
U_{sj}=\mathbf1_{\{j\ge s\}},\qquad\|U^{-1}\|_2\le2.
\label{eq:Jbase}
\end{equation}
The last inequality follows because $U^{-1}$ takes adjacent differences. Hence $\sigma_{\min}(\calJ)\ge1/(32\sqrt m)$. For the endpoint source, $w_1(t_\star)=1$ and $c_1^{(0)}=q_z(t_\star)/2\ge1/4$. Either perturbation bound retains $c_1\ge1/8$ for every phase bond, including the remote one.

The stated transfer and Jacobian constants apply to the ideal computable coefficients and times. A gauge-covariant rounding error with operator norm $\epsilon_H$ in the single-excitation sector changes each $c_s$ by at most $4t_\star\epsilon_H$. If this is the only implementation error, $\epsilon_H\le1/(128\sqrt m\,t_\star)$ retains $\sigma_{\min}(\calJ)\ge1/(64\sqrt m)$. Timing and preparation errors require separate allocations in the error budget.

\section{Quantum learning and classical hardness}
\label{app:learning-proof}
\label{app:pac}

The quantum learning proof combines uniform Hamiltonian compilation with regression on Fourier features. We account for noisy training labels and fresh feature measurements at prediction time, then prove the conditional hardness of classical prediction.

\subsection{Uniform compilation and resources}
\label{app:shared-gates}

The known local terms have fixed-dimensional support and polynomially computable matrix entries. For $H_N(\bm\beta)=\sum_{a=1}^{K_N}h_a(\bm\beta)$ with $\|h_a\|\le J_N$ and $|t|\le T_N$, the first-order product formula and finite-gate synthesis~\cite{Lloyd1996,Berry2007,LowChuang2019,Childs2021,DawsonNielsen2006} give an operator-norm approximation error at most $\varepsilon$ using
\begin{equation}
r=\left\lceil\frac{4K_N^2J_N^2T_N^2}{\varepsilon}\right\rceil+1
\label{eq:compiler-cost}
\end{equation}
steps. The product-formula error is at most $K_N^2J_N^2T_N^2/r$. Synthesizing each of the $K_Nr$ local exponentials to error $\varepsilon/(2K_Nr)$ adds at most $\varepsilon/2$. The gate count per evolution is therefore
\begin{equation}
G_{\rm sim}=O\!\left(K_Nr\,\operatorname{polylog}\frac{K_Nr}{\varepsilon}\right)
=\poly(N,T_N,1/\varepsilon).
\label{eq:shared-G}
\end{equation}
For feature precision $u$ in Eq.~\eqref{eq:pac-u}, choose $\varepsilon\le u/4$, so each simulated expectation value has systematic error at most $u/2$. Product-state preparation and the fixed local measurement add polynomial cost. The three-setting extraction below uses
\begin{align}
N_{\rm circuits}^{\rm(train)}&=3N_{\rm train}S_{\rm train},\label{eq:shared-train-cost}\\
N_{\rm circuits}^{\rm(test)}&=3S_{\rm pred}\label{eq:shared-test-cost}
\end{align}
executions, each with at most $G_{\rm sim}$ gates. Simulation circuits are compiled from the known Hamiltonian and settings, and the common classical labels supply all target-dependent information. The prediction cost includes fresh quantum feature measurements.

\subsection{Fourier extraction and regression}

For $s\le d$, the response at input $x=(z,s,\varphi,t)$ has only the frequencies $\bm0,\pm\bm u_s$. Evaluate the response at the prescribed phases $\theta_s=\beta_k=2\pi k/3$, with $k=0,1,2$ and all other coordinates zero. Since $\Phi_s=\beta_k$, a three-point Fourier inversion gives
\begin{equation}
\quad
\widehat F_x(r\bm u_s)=\frac13\sum_{k=0}^2 f_k\e^{-ir\beta_k},
\quad r\in\{-1,0,1\}.\quad
\label{eq:3dft}
\end{equation}

Here $f_k=F_x(\theta_s=\beta_k,\theta_{j\ne s}=0)$ is the simulated expectation value and $r$ is the scalar harmonic index. The phase $\beta_k$ is distinct from the response coefficient $\beta_{z,s,t}$. The coefficient $\widehat F_x(r\bm u_s)$ multiplies $\e^{ir\Phi_s}$ and includes the prepared source phase $\varphi$. For $s=m$, only the constant feature remains and is evaluated at zero phases.
Each $f_k$ is estimated by bounded local measurements. The shot counts needed for training and fresh prediction are given below.
Collect all Fourier features into
$\bm b(x)=(\widehat F_x(\bm\ell))_{\bm\ell\in\calS}$,
$\Mfeat=|\calS|\le2d+1$, and write
\begin{align}
F_x(\bm\theta_\star)
&=\Re\sum_{\bm\ell\in\calS}
b_{\bm\ell}(x)\,w_{\bm\ell}^\star,\nonumber\\
w_{\bm\ell}^\star
&=\e^{i\bm\ell\cdot\bm\theta_\star},
\qquad\norm{\bm w^\star}_2=\sqrt \Mfeat.
\label{eq:features}
\end{align}

Here $b_{\bm\ell}(x)=\widehat F_x(\bm\ell)$ is an input-dependent feature, and $\bm w^\star$ is the weight vector that represents the conditional mean. Estimate the features using Eq.~\eqref{eq:3dft} and minimize the unclipped empirical squared loss over $\|\bm w\|_2\le\sqrt\Mfeat$. Expressing the weights in real and imaginary parts gives a convex quadratically constrained quadratic program in $2\Mfeat$ real variables~\cite{BoydVandenberghe2004}. Each prediction uses fresh feature measurements and clips the resulting score to $[0,1]$.

\subsection{Learning with noisy labels}

To separate response error from label noise, set $g_{\bm w}(x)=\Re\sum_{\bm\ell\in\calS}b_{\bm\ell}(x)w_{\bm\ell}$ on $\|\bm w\|_2\le\sqrt\Mfeat$ and define
\[
\begin{aligned}
L(\bm w)&=\mathbb E_{(X,Y)}(g_{\bm w}(X)-Y)^2,\\
\sigma_Y^2&=\mathbb E_X\operatorname{Var}(Y\mid X).
\end{aligned}
\]

Here $(X,Y)$ is a fresh input--label pair, $L(\bm w)$ is the expected squared loss, and $\sigma_Y^2$ is the mean conditional label variance. Training pairs are independent, with $X_i\sim\mathcal D$, $Y_i\in[0,1]$, and $\mathbb E[Y_i\mid X_i]=c_{\bm\theta_\star}(X_i)$. Expanding the conditional square gives
\begin{equation}
L(\bm w)=\mathcal R_{\mathcal D}(g_{\bm w})+\sigma_Y^2.
\label{eq:noise-excess}
\end{equation}
The weights in Eq.~\eqref{eq:features} realize the conditional mean, so $L(\bm w^\star)=\sigma_Y^2$. The excess risk of the fitted vector $\widehat{\bm w}$ therefore equals its mean squared error relative to the conditional mean.

\subsubsection{Feature-estimation error}

Since $0\le F_x\le1$, every Fourier feature obeys $|b_{\bm\ell}(x)|\le1$, and $|g_{\bm w}(x)|\le\Mfeat$ on the weight ball. Choose the feature precision
\begin{equation}
\begin{aligned}
u&=\frac{\eta}{16\Mfeat(\Mfeat+2)},\qquad e=\Mfeat u,\\
\alpha&=2(\Mfeat+2)e=\frac\eta8.
\end{aligned}
\label{eq:pac-u}
\end{equation}
Here $u$ bounds each feature-coordinate error, $e=\Mfeat u$ bounds the induced prediction-score error, and $\alpha$ bounds the induced squared-loss error. For each training input $X_i$, extract a vector
$\widetilde{\bm b}_i$ by the three-setting inversion, using independent
shots of the simulation circuits. Require
$|\widetilde b_{i,\bm\ell}-b_{\bm\ell}(X_i)|\le u$
for all $i,\bm\ell$. The selection rule leaves at most three potentially nonzero coordinates
per input, with the remaining coordinates set to zero. A simulation systematic error of at most
$u/2$ in each of the three simulated expectation values, together with
at least
\begin{equation}
S_{\rm train}\ge\frac{2}{u^2}
\log\frac{12N_{\rm train}}{\delta}
\label{eq:pac-train-shots}
\end{equation}
independent bounded measurement shots per setting, ensures this event
with probability at least $1-\delta/2$ (Hoeffding's inequality~\cite{Hoeffding1963} and a union bound).
Define the estimated-feature score $\widetilde g_{i,\bm w}=\Re\sum_{\bm\ell\in\calS}\widetilde b_{i,\bm\ell}w_{\bm\ell}$. On that event, for every $\bm w$ in the ball and every $i$,
$|\widetilde g_{i,\bm w}-g_{\bm w}(X_i)|\le e$, by Cauchy--Schwarz.
Since $e\le1$ and $0\le Y_i\le1$,
\begin{align}
&\big|(\widetilde g_{i,\bm w}-Y_i)^2
 -(g_{\bm w}(X_i)-Y_i)^2\big|\nonumber\\
&\hspace{1em}\le e[2(\Mfeat+1)+e]\le\alpha.
\label{eq:pac-perturb}
\end{align}
Define the empirical losses
\[
\begin{aligned}
\widehat L_{\rm train}(\bm w)
&=\frac1{N_{\rm train}}\sum_{i=1}^{N_{\rm train}}
(g_{\bm w}(X_i)-Y_i)^2,\\
\widetilde L_{\rm train}(\bm w)
&=\frac1{N_{\rm train}}\sum_{i=1}^{N_{\rm train}}
(\widetilde g_{i,\bm w}-Y_i)^2.
\end{aligned}
\]
They use ideal and estimated features, respectively, and differ uniformly by at most $\alpha$.

\subsubsection{Uniform convergence and training risk}

The pointwise ideal-feature loss takes values in $[0,(\Mfeat+1)^2]$ and is $2(\Mfeat+1)\sqrt\Mfeat$-Lipschitz in the weights. A radius-$\rho$ net of the real $2\Mfeat$-dimensional parameter ball has at most $(1+2\sqrt\Mfeat/\rho)^{2\Mfeat}$ points. Set $\rho=\eta/[64(\Mfeat+1)\sqrt\Mfeat]$. Hoeffding's inequality~\cite{Hoeffding1963}, a union bound on the net, and the Lipschitz extension~\cite{CuckerSmale2002,BartlettMendelson2002,Bousquet2004} give
\begin{equation}
\sup_{\|\bm w\|_2\le\sqrt\Mfeat}
|\widehat L_{\rm train}(\bm w)-L(\bm w)|
\le\gamma_{\rm gen}=\frac\eta8
\label{eq:pac-uniform}
\end{equation}
with probability at least $1-\delta/2$, provided
\begin{equation}
\begin{multlined}
N_{\rm train}\ge\frac{128(\Mfeat+1)^4}{\eta^2}\\
{}\times\left[2\Mfeat\log\!\left(1+\frac{128\Mfeat(\Mfeat+1)}\eta\right)
+\log\frac4\delta\right].
\end{multlined}
\label{eq:pac--samples}
\end{equation}
The net deviation allowance is $\eta/16$, and the two extension errors sum to at most $\eta/16$. The symbol $\gamma_{\rm gen}$ is distinct from the hopping scale of the auxiliary chain.

Compute an approximate minimizer of the unclipped convex
quadratic program satisfying
$\widetilde L_{\rm train}(\widehat{\bm w})
\le\widetilde L_{\rm train}(\bm w^\star)+\eta/8$.
When feature estimation and uniform convergence both succeed, an event of probability at least $1-\delta$, the two empirical-loss errors and the two generalization errors give
\begin{equation}
\begin{aligned}
\mathcal R_{\mathcal D}(g_{\widehat{\bm w}})
&=L(\widehat{\bm w})-L(\bm w^\star)\\
&\le\eta/8+2\alpha+2\gamma_{\rm gen}
=\frac{5\eta}{8}.
\end{aligned}
\label{eq:pac-train-risk}
\end{equation}
The label variance cancels by Eq.~\eqref{eq:noise-excess}.
An approximate minimizer to additive objective precision $\eta/8$
can be found in polynomial time by standard convex optimization methods~\cite{BoydVandenberghe2004}.

\subsubsection{Prediction with fresh quantum measurements}

The training bound uses ideal features. For each new query, the protocol estimates them again and outputs
\[
h_{\widehat{\bm w}}(x;\omega)
=\Pi_{[0,1]}\!\left(
\Re\sum_{\bm\ell}\widetilde b_{\bm\ell}(x;\omega)
\widehat w_{\bm\ell}\right),
\]
where $\omega$ is its fresh shot randomness and
$\Pi_{[0,1]}$ is clipping to $[0,1]$.
Use simulation error at most $u/2$ and
$S_{\rm pred}\ge (2/u^2)\log(48/\eta)$ shots per setting.
For every $x$, the feature-error event then has probability at
least $1-\eta/8$. On success, clipping is nonexpansive and
$|h_{\widehat{\bm w}}-\Pi_{[0,1]}g_{\widehat{\bm w}}|
\le e$. Since $c_{\bm\theta_\star}(x)\in[0,1]$,
clipping does not increase squared error to the target. The
successful-shot squared error is therefore bounded by
$|g_{\widehat{\bm w}}(x)-c_{\bm\theta_\star}(x)|^2+2e$;
on failed shots it is at most $1$.
Conditioned on successful training, the randomized prediction risk obeys
\begin{align}
&\mathbb E_{X\sim\mathcal D,\omega}
|h_{\widehat{\bm w}}(X;\omega)-c_{\bm\theta_\star}(X)|^2
\nonumber\\
&\quad\le \frac{5\eta}{8}+2e+\frac\eta8
=\frac{3\eta}{4}+\frac{\eta}{8(\Mfeat+2)}
<\eta.
\label{eq:pac-pred-risk}
\end{align}

Sample size, compilation, circuit execution, convex optimization, and fresh per-query measurements all require polynomial resources, with bounds in Eqs.~\eqref{eq:shared-G}--\eqref{eq:shared-test-cost}. The resulting learner satisfies the distribution-free guarantee for deterministic or bounded conditionally unbiased labels.
\subsection{Conditional hardness of classical prediction}
\label{sec:hard}

The reduction from BQP circuit acceptance~\cite{BernsteinVazirani1997} uses the synchronized output gap. We assume polynomial classical resources for both training and evaluation of the output predictor.

Fix the target $\bm\theta_\star=\bm0$ and the family in Theorem~\ref{thm:unified}. For each promise input, use $x_z=(z,1,0,t_\star)$ with the same rational observation time for all instances of that input length. The reference response is $F^{(0)}_{x_z}(\bm0)=q_z(t_\star)$. Its perturbation is at most $2t_\star g_{\rm int}\le\varepsilon_{\rm int}$ for general couplings, or $2g^2t_\star^2\le\varepsilon_{\rm int}$ for the antisymmetric family. Thus both classes obey
\begin{equation}
\begin{gathered}
f_z:=F_{x_z}(\bm0),\qquad
f_z\ \begin{cases}
\ge f_{\rm Y},&z\in L_{\rm yes},\\
\le f_{\rm N},&z\in L_{\rm no},
\end{cases}\\
|f_z-\tau_{\rm pred}|>\Delta_{\rm test}.
\end{gathered}
\label{eq:gap}
\end{equation}

Here $f_z$ is the fixed-input response, with bounds $f_{\rm Y}=0.87025$ and $f_{\rm N}=0.62975$. The threshold is $\tau_{\rm pred}=3/4$, and $\Delta_{\rm test}=1/10$ is smaller than either threshold margin. These values follow from $\Delta_{\rm sync}=97/200$ and $\varepsilon_{\rm int}=10^{-3}$. For target prediction risk $\eta_{\rm hard}=10^{-3}$, thresholding then gives error at most $p_{\rm err}=1/10$ by Markov's inequality.
Fix the BQP input length $n$ and let $Z_n$ be the finite set of valid promise inputs.

For any distribution $\mathcal D_z$ on $Z_n$, draw independent training inputs $z\sim\mathcal D_z$ and, conditional on each input, a single-bit label $Y\sim\mathrm{Bernoulli}(f_z)$. Labels from different examples are drawn independently.
These single-bit labels satisfy the same bounded, conditionally
unbiased assumptions used by the quantum learner.

Suppose the hypothesized uniform randomized polynomial-time
classical learner works for every distribution and target,
and outputs a predictor with a polynomial-length classical description
that can be evaluated in classical polynomial time, with expected (over evaluation randomness) prediction risk at most $\eta=\eta_{\rm hard}$ on its successful training run.
Fix a successful training run, including its finite data and training coins. Averaging over evaluation randomness yields a fixed choice of prediction coins with no greater $\mathcal D_z$-averaged risk. This gives a nonuniform existence argument.
Let $L(z)$ be the promise answer bit, equal to one on $L_{\rm yes}$ and zero on $L_{\rm no}$, and let $C(z)$ be the Boolean circuit obtained by thresholding the predictor at $\tau_{\rm pred}$. The gap is at least
$\Delta_{\rm test}$, so Markov yields
\begin{equation}
\Pr_{z\sim\mathcal D_z}[C(z)\ne L(z)]
\le\eta/\Delta_{\rm test}^2=p_{\rm err}.
\end{equation}
The fixed choices may depend on $\mathcal D_z$, but the learner's polynomial resource bound gives a circuit-size bound independent of that distribution.
Define $\mathrm{err}(C,z)=\mathbf1_{\{C(z)\ne L(z)\}}$ and let $\mathcal C_{p(n)}$ contain Boolean circuits of size at most a fixed polynomial $p(n)$, independent of $\mathcal D_z$. Then
\begin{equation}
\forall\mathcal D_z\quad\exists C\in\mathcal C_{p(n)}:\quad
\mathbb E_{z\sim\mathcal D_z}\mathrm{err}(C,z)\le p_{\rm err}.
\end{equation}
The set of Boolean circuits $\mathcal C_{p(n)}$ is finite because
their descriptions have polynomially many bits. To obtain a pointwise error bound independent of $\mathcal D_z$, apply the finite von Neumann minimax theorem~\cite{vonNeumann1928} to the payoff $\mathrm{err}(C,z)\in\{0,1\}$.
There exists a distribution $\Pi_n$ on these circuits such that
\begin{equation}
\forall z\in Z_n:\quad
\Pr_{C\sim\Pi_n}[C(z)\ne L(z)]\le p_{\rm err}.
\end{equation}
Draw $k=cn$ independent circuits from $\Pi_n$, where $k$ is the ensemble size and $c>0$ is a constant, and take their majority. For each fixed $z$, Hoeffding bounds majority failure by
$\exp[-2k(1/2-p_{\rm err})^2]$. Choosing a fixed sufficiently large $c$
makes this below $2^{-n-1}$. A union bound over at most $2^n$ valid
inputs proves the existence of one fixed tuple of $O(n)$ circuits
whose majority is correct on every valid $z$.
Hardwire that tuple as polynomial-length nonuniform advice, using the same amplification-and-fixing principle as Adleman's nonuniform simulation argument~\cite{Adleman1978}.
The BQP-complete promise language then has polynomial-size classical
circuits, and uniform polynomial-time BQP reductions imply
$\BQP\subseteq\Ppoly$.

For the antisymmetric family, the full-response bound in Eq.~\eqref{eq:quad-response} preserves the classification gap. The reduction assumes a uniform polynomial-time learner that works for every input distribution and outputs a predictor with a polynomial-length classical description and classical polynomial-time evaluation.

\section{Uniform-XY chain}
\label{app:minimal}

For uniform hopping along a single chain, a coefficient associated with a bond next to the detector remains BQP-complete, while the phase sensitivity of a far-source response is factorially suppressed. This example separates coefficient hardness from the transport needed to retain a computational decision gap at a distant detector.

Uniform spin chains provide quantum communication channels~\cite{Bose2003}. The near-detector hard coefficient studied here uses uniform hopping in the weak-hopping regime $Jt_\star=1$, with $J=1/t_\star$. The suppression bound below applies to this scaling. Figure~\ref{fig:matched-transport} and Appendix~\ref{app:numeric} compare uniform and PST profiles at equal maximum bond strength over a common time interval.

The computational register and common observation time are those of Appendices~\ref{app:hqca} and~\ref{app:unified}. Here the auxiliary chain is a qubit chain, with coherence between the vacuum and a single excitation. This choice gives local dimension $20$, compared with $30$ for the construction with two occupied internal states. We derive the near-detector transfer amplitude and far-source suppression below. The learning argument follows Appendix~\ref{app:learning-proof}.

In this appendix, $m=m_{\rm ch}$ denotes the qubit-chain length and $d_{\rm ch}=m-1$ its number of phase bonds. These lengths are distinct from the ladder parameters. The phase vector lies in $\T^{d_{\rm ch}}$, and $\bm u_s$ and $\bm e_j$ are the corresponding indicator and coordinate vectors. The basis state $\ket j$ has one excitation in the auxiliary chain at position $j$, and $A_m=\sum_{j=1}^{m-1}(\ket j\bra{j+1}+\mathrm{h.c.})$ is the unit-hopping adjacency. We abbreviate the response coefficients as $\widehat F_{z,s,t}(\bm\ell)=\widehat F^{\rm ch}_{z,s}(\bm\ell,t)$.
\begin{theorem}[BQP-complete coefficient in a uniform chain]
\label{thm:minimal}

For every BQP-complete promise problem, there is a uniformly constructible, time-independent family $H_n^{\rm ch}(\bm\theta)$ on a single nearest-neighbor chain of $L=24M$ sites, with local dimension at most $20$ and bounded local terms. It uses the ten-state computational register and efficiently computable, instance-independent observation time $t_\star=\poly(n)$ of Theorem~\ref{thm:unified}. An auxiliary qubit chain occupies $m_{\rm ch}=M+1$ consecutive computational sites ending at $a$. Its $d_{\rm ch}=M$ XY bonds carry independent phases and have uniform strength $J=1/t_\star$. A nonzero on-site coupling between the occupation number and the data spin has strength $g=1/(2000t_\star)$. Inputs are product states, and the readout is a fixed single-site measurement effect.
For each source $s$, the Fourier support at every
evolution time is contained in
$\{\bm0,\pm\bm u_s\}$, where
$(\bm u_s)_j=\mathbf1_{\{j\ge s\}}$ and the common Fourier frequency set has size
at most $2d_{\rm ch}+1$. At the synchronized observation time $t_\star$, the coefficient
for source $s=d_{\rm ch}$ satisfies
\begin{equation}
\left|\widehat F_{z,d_{\rm ch},t_\star}(\bm e_{d_{\rm ch}})
-\frac{3+s_z(t_\star)}{16}R_m\right|\le 10^{-3},
\qquad R_m>\frac35,
\label{eq:minimal-hard}
\end{equation}
where $R_m=i\langle m|\e^{-iA_m}|m-1\rangle>0$ is an efficiently computable
phase-adjusted endpoint transfer amplitude. Estimating the real part of this specified
coefficient to additive accuracy $1/200$ is BQP-complete, and its
magnitude exceeds $0.074$.

By contrast, for a source at the opposite endpoint $s=1$, at the
same time $t_\star$ the Fourier amplitude obeys
\begin{equation}
\left|\widehat F_{z,1,t_\star}(\bm u_1)\right|
\le\frac{\e^2}{4}\frac{2^{d_{\rm ch}}}{d_{\rm ch}!}.
\label{eq:minimal-remote}
\end{equation}
The same family admits a polynomial-resource quantum-gate PAC learner.
A polynomial-time classical learner given the same classical training data
and Hamiltonian description, with classically evaluable output and a
sufficiently small distribution-free prediction risk, would imply
$\BQP\subseteq\Ppoly$.
\end{theorem}

\subsection{Local realization and Fourier support}

Use the padded HQCA of Appendix~\ref{app:hqca}, with $M=wk$, $L=24M$, and output $a=22M+w$. The deterministic time search gives the same $t_\star$ and computational output margin $\Delta_{\rm sync}=97/200$ as in Appendix~\ref{app:unified}.

Place auxiliary qubits at $p_j=a-m+j$ for $1\le j\le m$, where $m=M+1$ and $d_{\rm ch}=M$. All positions lie in the computational chain and $p_m=a$. Each overlapping site has dimension $10\times2=20$. The Hamiltonian is
\begin{align}
H_n^{\rm ch}(\bm\theta)
&=H_{{\rm Q},n}+H_{\rm XY}(\bm\theta)+V_{\rm ch},
\label{eq:minimal-H}\\
H_{\rm XY}(\bm\theta)
&=J\sum_{j=1}^{d_{\rm ch}}
\bigl(\e^{i\theta_j}\sigma_j^+\sigma_{j+1}^-
+\e^{-i\theta_j}\sigma_j^-\sigma_{j+1}^+\bigr),
\label{eq:minimal-XY}\\
V_{\rm ch}&=g\sum_{j=1}^m n_j\otimes Z_{p_j}^{\rm data}.
\label{eq:minimal-V}
\end{align}

Here $J=t_\star^{-1}$, $g=(2000t_\star)^{-1}$, and $n_j=\ket1\bra1_{p_j}$ is the local occupation in the auxiliary chain. The operators $\sigma_j^+=\ket1\bra0_{p_j}$ and $\sigma_j^-=\ket0\bra1_{p_j}$ create and annihilate an excitation in the auxiliary chain. The operator $Z_{p_j}^{\rm data}$ is Pauli $Z$ on the data qubit. The interaction $V_{\rm ch}$ acts between registers at the same physical site. All other terms are nearest neighbor. Excitation number is conserved, while $V_{\rm ch}$ generally does not commute with hopping.

Prepare
\begin{equation}
\ket{\psi_s}=\frac{\ket{\mathrm{vac}}+\ket{s}}{\sqrt2},
\qquad
\rho_{z,s}=\rho_z^{\rm Q}\otimes\ket{\psi_s}\bra{\psi_s}.
\label{eq:minimal-source}
\end{equation}

Here $\ket{\mathrm{vac}}=\ket0^{\otimes m}$ and $\ket s$ has one excitation at $p_s$, with $1\le s\le d_{\rm ch}$. Only the source qubit is in $\ket+=(\ket0+\ket1)/\sqrt2$, so the preparation is a product state across sites. The detector compares the arriving excitation with the vacuum using
\begin{equation}
\begin{aligned}
O_A^{\rm ch}&=B_a\otimes P_{-y,a},\\
P_{-y,a}&=\frac{I-Y_a}{2},\qquad
0\le O_A^{\rm ch}\le I.
\end{aligned}
\label{eq:minimal-readout}
\end{equation}

Here $B_a=(3I+O_a^{\rm Q})/4$, $Y_a$ is Pauli $Y$ on the endpoint transport qubit, and $P_{-y,a}$ projects onto its eigenvalue $-1$ state. The unconditional readout includes the vacuum contribution and requires a phase reference between different excitation-number sectors~\cite{BartlettRudolphSpekkens2007}.

The phase dependence again follows from gauge covariance:
\begin{equation}
H_n^{\rm ch}(\bm\theta)
=D_{\rm ch}(\bm\theta)H_n^{\rm ch}(\bm0)
D_{\rm ch}(\bm\theta)^\dagger,
\quad [O_A^{\rm ch},D_{\rm ch}]=0.
\label{eq:minimal-gauge}
\end{equation}

Here $D_{\rm ch}(\bm\theta)=\prod_{j=1}^m\e^{i\Phi_jn_j}$, with $\Phi_j=\sum_{k=j}^{d_{\rm ch}}\theta_k$ and $\Phi_m=0$. The rotation fixes the vacuum and multiplies $\ket s$ by $\e^{i\Phi_s}$. Expanding the transformed preparation gives
\begin{equation}
F_{z,s}^{\rm ch}(\bm\theta,t)
=A_{z,s,t}+C_{z,s,t}\e^{i\Phi_s(\bm\theta)}
+\overline{C_{z,s,t}}\e^{-i\Phi_s(\bm\theta)}.
\label{eq:minimal-three}
\end{equation}

Here $F_{z,s}^{\rm ch}$ is the evolved expectation of $O_A^{\rm ch}$ from $\rho_{z,s}$. The background $A_{z,s,t}\in\R$ and coherence coefficient $C_{z,s,t}\in\C$ are phase independent. Thus the Fourier support of the response is contained in $\{\bm0,\pm\bm u_s\}$, with a common Fourier frequency set of size $2d_{\rm ch}+1$, at every time and interaction strength in this gauge-covariant class.
\subsection{BQP-complete coefficient near the detector}

Transfer across the last bond of the transport chain is determined by
\begin{equation}
R_m=i\bra m\e^{-iA_m}\ket{m-1}.
\label{eq:minimal-R}
\end{equation}

Here $A_m$ is the unit-hopping adjacency defined above. Since $Jt_\star=1$, $R_m$ is the phase-adjusted reference transfer amplitude from $s=m-1$ to the endpoint. Bipartiteness gives the real odd-walk series $R_m=1-c_3/3!+c_5/5!-\cdots$, where $c_k=\bra mA_m^k\ket{m-1}\ge0$ counts length-$k$ walks. Using $c_3\le2$ and $c_k\le2^k$,
\begin{equation}
\begin{split}
R_m&\ge1-\frac{2}{3!}
-\sum_{\substack{k\ge7\\k\,{\rm odd}}}\frac{2^k}{k!}\\
&\ge1-\frac13-\frac{16}{595}>\frac35
\qquad (m\ge2).
\end{split}
\label{eq:minimal-Rbound}
\end{equation}

The negative-tail bound $16/595$ follows by majorizing odd terms starting at $k=7$, whose successive ratio is at most $1/18$. The known amplitude $R_m$ is efficiently computable by finite-matrix exponentiation or the path spectrum.

For source $s=d_{\rm ch}=m-1$, the endpoint reference amplitude is $-iR_m\e^{-i\theta_{d_{\rm ch}}}$. Interference with the vacuum under $P_{-y,a}$ gives
\begin{equation}
F_{z,d_{\rm ch}}^{\rm ch,(0)}(\bm\theta,t_\star)
=\frac{3+s_z(t_\star)}{8}
\left(1+R_m\cos\theta_{d_{\rm ch}}\right).
\label{eq:minimal-ref}
\end{equation}

The coefficient at $+\bm e_{d_{\rm ch}}$ is therefore $(3+s_z(t_\star))R_m/16$. To bound the interaction correction, restrict to the invariant vacuum and one-excitation sectors. The occupation projectors have disjoint supports in the latter, giving
\begin{equation}
\norm{P_{\le1}V_{\rm ch}P_{\le1}}=g,
\quad
\left|F_{z,s}^{\rm ch}-F_{z,s}^{\rm ch,(0)}\right|
\le2|t|g.
\label{eq:minimal-duhamel}
\end{equation}

Here $P_{\le1}$ projects onto the combined vacuum and one-excitation space. Integrating the response difference against a normalized Fourier character bounds the coefficient correction by $2|t|g$. At $t_\star$, this equals $10^{-3}$ and proves Eq.~\eqref{eq:minimal-hard}. The real coefficient obeys
\begin{equation}
\begin{array}{ll}
\mathrm{YES}:&\Re\widehat F(\bm e_{d_{\rm ch}})
\ge\tau_m+\Gamma,\\
\mathrm{NO}:&\Re\widehat F(\bm e_{d_{\rm ch}})
\le\tau_m-\Gamma,
\end{array}
\quad
\Gamma>\frac{11}{640}>\frac1{200}.
\label{eq:minimal-gap}
\end{equation}

Here $\tau_m=3R_m/16$ is the decision threshold and $\Gamma=\Delta_{\rm sync}R_m/16-10^{-3}$ is a lower bound on its margin. The inequality uses $R_m>3/5$ and $\Delta_{\rm sync}=97/200$. Also, $B_a\ge I/2$ gives a reference coefficient at least $R_m/8>3/40$, so the coupled coefficient has magnitude greater than $0.074$. Accuracy $1/200$ therefore resolves BQP circuit acceptance. Conversely, the support restriction to $\{\bm0,\pm\bm u_s\}$ permits three-point inversion at $\theta_s=0,2\pi/3,4\pi/3$. Uniform simulation and bounded single-site measurements give polynomial quantum cost, proving BQP membership.
\subsection{Distant-source suppression and learning}

For a source at distance $r=m-s$, use the interaction picture with respect to $H_{{\rm Q},n}+V_{\rm ch}$, which preserves the excitation's position. Each XY insertion moves the excitation by at most one site, so the source-to-detector Dyson series starts at order $r$. Since the one-excitation hopping norm is at most $2J$ and $Jt_\star=1$,
\begin{equation}
\norm{\bra m\e^{-it_\star H_n^{\rm ch}(\bm0)}\ket s}
\le\sum_{k=r}^{\infty}\frac{2^k}{k!}
\le \e^2\frac{2^r}{r!}.
\label{eq:minimal-dyson}
\end{equation}

The bracket is an operator on the computational space. Phase dependence comes only from coherence between the vacuum and an excitation at the detector, whose Fourier coefficient is at most one quarter of this norm because $\|B_a\|\le1$. Setting $s=1$ and $r=d_{\rm ch}$ proves Eq.~\eqref{eq:minimal-remote}. The three-frequency form~\eqref{eq:minimal-three} also converts this coefficient estimate into a susceptibility bound. For the far source, $\bm u_1$ contains every phase bond, so at every phase setting
\begin{equation}
\left|\partial_{\theta_j}F_{z,1}^{\rm ch}(\bm\theta,t_\star)\right|
\le2\left|\widehat F_{z,1,t_\star}(\bm u_1)\right|
\le\frac{\e^2}{2}\frac{2^{d_{\rm ch}}}{d_{\rm ch}!}.
\label{eq:minimal-remote-slope}
\end{equation}
At $Jt_\star=1$, coefficient hardness near the detector coexists with a factorially suppressed distant-source susceptibility. PST in the main construction preserves a resolvable computational gap as the source--detector separation grows.

The same sparse-feature learner applies to this chain. Given inputs $x=(z,s,t)$ and bounded unbiased labels with means $F_{z,s}^{\rm ch}(\bm\theta_\star,t)$, extract at most three features by simulation at the prescribed phase settings and Fourier inversion. Constrained least squares and fresh feature measurements at prediction time then give polynomial sample, training, and prediction resources. The proof in Appendix~\ref{app:learning-proof} applies with $\Mfeat\le2d_{\rm ch}+1$.

For the classical reduction, fix $\bm\theta_\star=\bm0$, $s=d_{\rm ch}$, and $t=t_\star$. The reference response and interaction bound give
\begin{equation}
\begin{gathered}
 f_z\ge q_m(3+\Delta_{\rm sync})/4-10^{-3}
 \quad(\mathrm{YES}),\\
 f_z\le q_m(3-\Delta_{\rm sync})/4+10^{-3}
 \quad(\mathrm{NO}),\\
 |f_z-3q_m/4|\ge
 \Delta_{\rm sync}q_m/4-10^{-3}>0.096.
\end{gathered}
\label{eq:minimal-pac-gap}
\end{equation}

Here $q_m=(1+R_m)/2>4/5$ is the known interference factor and $f_z=F_{z,d_{\rm ch}}^{\rm ch}(\bm0,t_\star)$. With margin $\Delta_{\rm test}=0.09$ and target prediction risk $\eta_{\rm hard}=10^{-3}$, thresholding a classically evaluable predictor at $3q_m/4$ gives error at most $10/81<1/2$. The finite-minimax and majority argument of Appendix~\ref{app:pac} then implies $\BQP\subseteq\Ppoly$ if a uniform efficient classical learner achieves this guarantee for every input distribution.

\section{Fourier leakage under gauge-breaking perturbations}
\label{app:universal-leak}

The sparse response and learning protocol in Secs.~\ref{sec:transport} and~\ref{sec:learning} rely on gauge covariance. The rung interactions in Sec.~\ref{sec:stability-main} change the Fourier coefficients within the same frequency set. Gauge-breaking perturbations can also populate forbidden frequencies. We bound the resulting approximation error and its contribution to prediction risk.

\subsection{Response error and approximate Fourier support}

Let $H^{\rm cov}(\bm\theta)$ be a known gauge-covariant reference family, including the allowed couplings between the computational register and auxiliary chain, and perturb it by a Hermitian $E(\bm\theta)$ such that
\[
\widetilde H(\bm\theta)=H^{\rm cov}(\bm\theta)+E(\bm\theta),\qquad
\sup_{\bm\theta}\|E(\bm\theta)\|\le\epsilon.
\]
Use the same preparations and fixed effect $O_A$ for both dynamics. For an input $x$ with evolution time $t$, Duhamel's formula gives a propagator difference of norm at most $|t|\epsilon$, hence
\begin{equation}
\sup_{\bm\theta}|\widetilde F_x(\bm\theta)-F_x^{\rm cov}(\bm\theta)|
\le2|t|\epsilon\|O_A\|.
\label{eq:leak-response}
\end{equation}
Here $\epsilon$ bounds the norm of the full perturbation. The bound also applies within any common invariant sector of the two evolutions.

Let $\Pi_{\calS}$ be the orthogonal projection onto the Fourier frequencies in Eq.~\eqref{eq:support}, with normalized phase measure $d^d\bm\theta/(2\pi)^d$. Since $(I-\Pi_{\calS})F_x^{\rm cov}=0$, Parseval's identity and Eq.~\eqref{eq:leak-response} imply
\begin{equation}
\begin{aligned}
\mathcal L_{\calS}(t)
&:=\left(\sum_{\bm\ell\notin\calS}
|\widehat{\widetilde F}_x(\bm\ell)|^2\right)^{1/2}\\
&=\|(I-\Pi_{\calS})\widetilde F_x\|_{L_2}
\le2|t|\epsilon\|O_A\|.
\end{aligned}
\label{eq:leak-general}
\end{equation}
The leakage is the best phase-averaged approximation error achievable with the reference frequency set:
\begin{equation}
\inf_{f:\,\supp\widehat f\subseteq\calS}
\|\widetilde F_x-f\|_{L_2}=\mathcal L_{\calS}(t).
\label{eq:leak-approximation}
\end{equation}
Thus weak gauge breaking gives controlled approximate sparsity on a finite observation interval. If the perturbation preserves gauge covariance, the forbidden coefficients are zero at every coupling strength.

\subsection{Locality of the leakage}

Suppose $E(\bm\theta)$ is supported in a fixed bounded region $Y$ at distance $r=\operatorname{dist}(A,Y)$ from the detector, and the reference family satisfies the uniform locality assumptions of Appendix~\ref{app:universal-lr}. For $t\ge0$, the difference of the Heisenberg evolutions satisfies
\[
\begin{aligned}
\widetilde\tau_t(O_A)-\tau_t^{\rm cov}(O_A)
={}&i\int_0^t\widetilde\tau_s
\big([E,\tau_{t-s}^{\rm cov}(O_A)]\big)\,ds.
\end{aligned}
\]
The reference Lieb--Robinson bound therefore controls the response difference uniformly in the phases. Projecting onto the forbidden frequencies gives
\begin{equation}
\mathcal L_{\calS}(t)\le
\frac{C_{\rm LR}\epsilon\|O_A\|}{\mu v}
\e^{-\mu(r-vt)}.
\label{eq:leak-lightcone}
\end{equation}
Taking the smaller bound in Eqs.~\eqref{eq:leak-general} and~\eqref{eq:leak-lightcone} gives an exponentially suppressed leakage from remote perturbations before their influence reaches the detector. The propagation bound therefore constrains the effect of remote errors on the Fourier representation.

\subsection{Prediction using reference features}

The uniform estimate~\eqref{eq:leak-response} gives a prediction-risk bound at a fixed unknown target phase. Suppose the bounded, conditionally unbiased labels have mean $\widetilde c(x)=\widetilde F_x(\bm\theta_\star)$, and all observation times obey $|t|\le T$. Use the known reference features $b_{\bm\ell}^{\rm cov}(x)=\widehat F_x^{\rm cov}(\bm\ell)$ in the algorithm of Appendix~\ref{app:learning-proof}, using circuits for the reference Hamiltonian. The perturbation $E$ enters through the labels. The phase weights $w_{\bm\ell}^\star=\e^{i\bm\ell\cdot\bm\theta_\star}$ realize the reference response, giving the mean squared error of the reference response
\begin{equation}
\begin{aligned}
A_{\mathcal D}
&:=\mathbb E_{x\sim\mathcal D}
|F_x^{\rm cov}(\bm\theta_\star)-\widetilde c(x)|^2\\
&\le(2T\epsilon\|O_A\|)^2.
\end{aligned}
\label{eq:leak-comparator}
\end{equation}
The label-variance identity~\eqref{eq:noise-excess} still holds with $\widetilde c$ as the conditional mean. Repeating the empirical-loss comparison~\eqref{eq:pac-train-risk} therefore gives training risk at most $A_{\mathcal D}+5\eta/8$ instead of $5\eta/8$. The fresh-measurement and clipping argument~\eqref{eq:pac-pred-risk} adds less than $3\eta/8$. With probability at least $1-\delta$ over training,
\begin{equation}
\begin{aligned}
\widetilde{\mathcal R}_{\mathcal D}(h)
&:=\mathbb E_{x\sim\mathcal D,\omega}
|h(x;\omega)-\widetilde c(x)|^2\\
&\le A_{\mathcal D}+\eta
\le(2T\epsilon\|O_A\|)^2+\eta.
\end{aligned}
\label{eq:leak-prediction-risk}
\end{equation}
The resource bounds remain polynomial in the same parameters as in Appendix~\ref{app:learning-proof}. The additional term is an upper bound on the contribution from model mismatch when using the reference features. Because the target phase is fixed and unknown, the proof uses the uniform response bound.

Fourier-tail bounds control the $L_2$ error over phases. Equation~\eqref{eq:leak-response} bounds the response at each phase setting. Bounds on phase derivatives require further restrictions on the phase dependence of $E$. The susceptibility and Jacobian results in the main text apply to the stated gauge-covariant interaction classes.

\section{Finite-size verification}
\label{app:numeric}

Figures~\ref{fig:stability}(b,c), \ref{fig:matched-transport}, and \ref{fig:fixed-coupling-sizes} use the finite models specified below to calculate coupled responses and excitation transport. Figures~\ref{fig:ballistic}, \ref{fig:stability}(a), and \ref{fig:identification-learning} use analytical expressions and bounds. BQP completeness and the common HQCA output gap are proved in Appendices~\ref{app:hqca} and~\ref{app:unified}.

\subsection{Transport comparison at matched strength and distance}

\begin{figure*}[t]
\centering
\includegraphics[width=\textwidth]{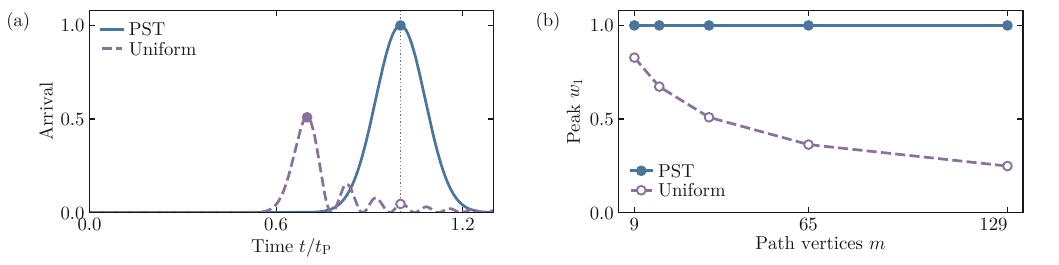}
\caption{Uncoupled PST and uniform transport at equal path length and maximum bond strength, matched using Eq.~\eqref{eq:matched-uniform}, with $\gamma=1.4$. Both profiles use the observation window $[0,1.3t_{\rm P}]$, where $t_{\rm P}=\pi m/(2\gamma)$. (a) Arrival probabilities at $m=33$. Filled circles mark each profile's sampled maximum; the open purple circle marks uniform arrival at $t_{\rm P}$. The dotted vertical line marks $t_{\rm P}$. (b) Peak probabilities in the common window, with filled blue and open purple circles for PST and uniform profiles, respectively. For the sampled sizes, PST maintains unit peak arrival within this observation window, while the uniform-chain peak decreases with path length.}
\label{fig:matched-transport}
\end{figure*}

For these uncoupled calculations, use $\gamma=1.4$ and $t_{\rm P}=\pi m/(2\gamma)$. For each length, compare the phase-free PST hopping matrix with the uniform matrix whose bond amplitude is
\begin{equation}
 J_{\rm uni}(m)=\max_{1\le j<m}J_j^{\rm geo}
 =\frac{\gamma}{m}\sqrt{\left\lfloor\frac{m^2}{4}\right\rfloor}.
\label{eq:matched-uniform}
\end{equation}
Both chains use the same initial endpoint excitation, number of sites, maximum bond strength, detector, and observation-time interval $[0,1.3t_{\rm P}]$. The uncoupled endpoint probabilities, as shown in Fig.~\ref{fig:matched-transport}(a), are evaluated at $m=33$ from the tridiagonal hopping spectrum. The uniform chain's grid maximum is $0.5091$ near $0.6995t_{\rm P}$, while its value at $t_{\rm P}$ is $0.0460$. The PST probability at $t_{\rm P}$ is one. To account for the different arrival times, Fig.~\ref{fig:matched-transport}(b) shows the maxima on a common grid of $2601$ time points for $m=9,17,33,65,129$. The uniform maxima are $0.8281,0.6730,0.5091,0.3633,0.2485$, respectively. The comparison covers these system sizes, hopping profiles, and observation interval. The analytical comparison in Appendix~\ref{app:minimal} uses the scaling $Jt_\star=1$, for which a hard near-detector coefficient coexists with a suppressed remote susceptibility.

\subsection{Three-spin register and coupled susceptibility}

For this finite test, replace the HQCA register by three data spins on bottom-row sites $u=1,2,3$, with
\begin{equation}
\begin{aligned}
H_{\rm Q}^{\rm toy}
 &=\sum_{u=1}^3(h_u^xX_u^{\rm data}+h_u^zZ_u^{\rm data})\\
 &\quad+0.19Z_1^{\rm data}Z_2^{\rm data}
       +0.13X_2^{\rm data}X_3^{\rm data},\\
\bm h^x&=(0.43,0.37,0.31),\\
\bm h^z&=(0.27,0.17,-0.23).
\end{aligned}
\label{eq:toy-computer}
\end{equation}
Here $X_u^{\rm data}$ and $Z_u^{\rm data}$ are Pauli operators. Set $L=3$, output column $a=2$, and $\ell=(m+1)/2$ for the odd lengths considered. The auxiliary chain follows the ladder path of Sec.~\ref{sec:unified}. Its first three vertices lie above the respective data spins, so all rung interactions remain local. Use $\gamma=1.4$, initial data state $\ket{000}$, source $s=1$, and computational effect $B^{\rm toy}=(3I+Z_2^{\rm data})/4$. The three-spin dynamics are sampled at $t_{\rm P}=\pi m/(2\gamma)$, the uncoupled PST time.

For the family with unequal couplings, take
\begin{equation}
 W_{u,0}^{\rm toy}=\frac7{20}gZ_u^{\rm data},\qquad
 W_{u,1}^{\rm toy}=gZ_u^{\rm data},\quad 1\le u\le3.
\label{eq:toy-unequal}
\end{equation}
For the antisymmetric family, take $W_{u,0}^{\rm toy}=-gZ_u^{\rm data}$ and $W_{u,1}^{\rm toy}=gZ_u^{\rm data}$. In both families, the interaction norm in the single-excitation sector is $|g|$. With tensor order $\C^m\otimes\C^8$, let $\mathcal H_r^{\rm toy}(g)$ be the Hamiltonian restricted to internal state $r$ and define
\begin{equation}
\begin{aligned}
 \ket{\xi_r(g)}&=(\bra m\otimes I)
 \e^{-it_{\rm P}\mathcal H_r^{\rm toy}(g)}
 (\ket1\otimes\ket{000}),\\
 \chi_{\rm toy}(g)&=\frac12\Re
 \bra{\xi_0(g)}B^{\rm toy}\ket{\xi_1(g)}.
\end{aligned}
\label{eq:toy-susceptibility}
\end{equation}
Equation~\eqref{eq:toy-susceptibility} follows from the same parameter-shift rule as Eq.~\eqref{eq:two-reference-readout}. Its uncoupled value is $\chi_{\rm toy}^{(0)}=q^{\rm toy}(t_{\rm P})/2$, where $q^{\rm toy}(t)$ is the expectation of $B^{\rm toy}$ in the state $\e^{-itH_{\rm Q}^{\rm toy}}\ket{000}$.

\begin{figure}[t]
\centering
\includegraphics[width=\columnwidth]{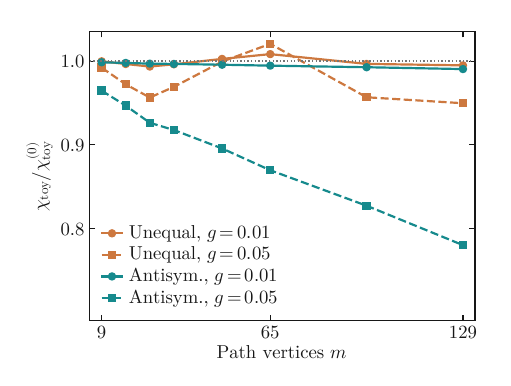}
\caption{Finite-size responses of the three-spin model in Appendix~\ref{app:numeric} at fixed couplings $g=0.01,0.05$, with $\gamma=1.4$ and observation time $t_{\rm P}=\pi m/(2\gamma)$. Susceptibilities are normalized by the uncoupled value at the same size and time; the dotted line denotes one. In these finite-size tests, antisymmetric coupling does not uniformly reduce the susceptibility error at fixed coupling.}
\label{fig:fixed-coupling-sizes}
\end{figure}

The scripts evaluate sparse matrix-exponential actions using SciPy's \texttt{expm\_multiply}, which implements the algorithm of Al-Mohy and Higham~\cite{AlMohyHigham2011}, without sampling noise. Figure~\ref{fig:stability}(b) uses $81$ equally spaced values of $gt_{\rm P}$ in $[-1,1]$ at $m=33$. Figure~\ref{fig:stability}(c) fits a line to the logarithm of the absolute error against $\log(|g|t_{\rm P})$, using $11$ geometrically spaced positive values between $10^{-4}$ and $10^{-2}$. The fitted slopes are $1.0008$ for the family with unequal couplings and $2.0000$ for the antisymmetric family. Figure~\ref{fig:fixed-coupling-sizes} uses fixed $g=0.01,0.05$ at $m=9,17,25,33,49,65,97,129$. Normalization by the uncoupled susceptibility at each length removes the variation in the uncoupled computational expectation.

\subsection{Checks and numerical accuracy}

The scripts compare sparse propagation with dense matrix exponentiation, check norm conservation, and compare the susceptibility from the parameter-shift rule with a finite difference obtained by changing a hopping phase and with a three-point Fourier inversion. They also check evenness in the coupling under exchange of the internal states, the uncoupled transfer law, the binomial peak bound, and the response Jacobian for every source at $m=9,17,33$ with the respective allowed coupling strengths. In these runs, norm errors are below $10^{-13}$, dense--sparse differences are below $10^{-13}$, and the central phase difference with step $10^{-4}$ differs from the value from the parameter-shift rule by less than $10^{-8}$. The Fourier and exchange-symmetry checks agree within $10^{-13}$. The tested cases satisfy the bounds in Eqs.~\eqref{eq:pst-uniform} and~\eqref{eq:chi-jacobian}. The numerical archive includes the calculated values, fitting inputs, check summaries, and package versions.

\bibliography{quantum_peierls_references}
@article{NagajWocjan2008,
  author={Nagaj, Daniel and Wocjan, Pawel},
  title={Hamiltonian quantum cellular automata in one dimension},
  journal={Phys. Rev. A},
  volume={78},
  pages={032311},
  year={2008},
  doi={10.1103/PhysRevA.78.032311},
  eprint={0802.0886},
  archivePrefix={arXiv},
  primaryClass={quant-ph}
}

@article{Christandl2004,
  author={Christandl, Matthias and Datta, Nilanjana and Ekert, Artur and Landahl, Andrew J.},
  title={Perfect state transfer in quantum spin networks},
  journal={Phys. Rev. Lett.},
  volume={92},
  pages={187902},
  year={2004},
  doi={10.1103/PhysRevLett.92.187902},
  eprint={quant-ph/0309131},
  archivePrefix={arXiv}
}

@article{Vinet2012,
  author={Vinet, Luc and Zhedanov, Alexei},
  title={How to construct spin chains with perfect state transfer},
  journal={Phys. Rev. A},
  volume={85},
  pages={012323},
  year={2012},
  doi={10.1103/PhysRevA.85.012323}
}

@article{NachtergaeleSims2006,
  author={Nachtergaele, Bruno and Sims, Robert},
  title={Lieb--{R}obinson bounds and the exponential clustering theorem},
  journal={Commun. Math. Phys.},
  volume={265},
  number={1},
  pages={119--130},
  year={2006},
  doi={10.1007/s00220-006-1556-1},
  eprint={math-ph/0506030},
  archivePrefix={arXiv}
}

@misc{Barthe2025,
  author={Barthe, Alice and Yaghubi Rad, Mahtab and Grossi, Michele and Dunjko, Vedran},
  title={Quantum advantage in learning quantum dynamics via {Fourier} coefficient extraction},
  year={2025},
  eprint={2506.17089},
  archivePrefix={arXiv},
  primaryClass={quant-ph},
  url={https://arxiv.org/abs/2506.17089}
}

@article{Peierls1933,
 author={Peierls, Rudolf}, title={{Zur Theorie des Diamagnetismus von Leitungselektronen}},
 journal={Zeitschrift f{\"u}r Physik}, volume={80}, pages={763--791}, year={1933},
 doi={10.1007/BF01342591}
}

@article{Dalibard2011,
 author={Dalibard, Jean and Gerbier, Fabrice and Juzeli{\=u}nas, Gediminas and {\"O}hberg, Patrik},
 title={Colloquium: Artificial Gauge Potentials for Neutral Atoms},
 journal={Reviews of Modern Physics}, volume={83}, pages={1523--1543}, year={2011},
 doi={10.1103/RevModPhys.83.1523}
}

@article{Bose2003,
 author={Bose, Sougato}, title={Quantum Communication through an Unmodulated Spin Chain},
 journal={Phys. Rev. Lett.}, volume={91}, pages={207901}, year={2003},
 doi={10.1103/PhysRevLett.91.207901}
}

@article{Albanese2004,
 author={Albanese, Claudio and Christandl, Matthias and Datta, Nilanjana and Ekert, Artur},
 title={Mirror Inversion of Quantum States in Linear Registers},
 journal={Phys. Rev. Lett.}, volume={93}, pages={230502}, year={2004},
 doi={10.1103/PhysRevLett.93.230502}
}

@article{Yung2006,
 author={Yung, Man-Hong}, title={Quantum Speed Limit for Perfect State Transfer in One Dimension},
 journal={Phys. Rev. A}, volume={74}, pages={030303}, year={2006},
 doi={10.1103/PhysRevA.74.030303}
}

@article{Lloyd1996,
 author={Lloyd, Seth}, title={Universal Quantum Simulators},
 journal={Science}, volume={273}, pages={1073--1078}, year={1996},
 doi={10.1126/science.273.5278.1073}
}

@article{Childs2009,
 author={Childs, Andrew M.}, title={Universal Computation by Quantum Walk},
 journal={Phys. Rev. Lett.}, volume={102}, pages={180501}, year={2009},
 doi={10.1103/PhysRevLett.102.180501}
}

@article{BernsteinVazirani1997,
 author={Bernstein, Ethan and Vazirani, Umesh}, title={Quantum Complexity Theory},
 journal={SIAM J. Comput.}, volume={26}, pages={1411--1473}, year={1997},
 doi={10.1137/S0097539796300921}
}

@article{Berry2007,
 author={Berry, Dominic W. and Ahokas, Graeme and Cleve, Richard and Sanders, Barry C.},
 title={Efficient Quantum Algorithms for Simulating Sparse {Hamiltonians}},
 journal={Commun. Math. Phys.}, volume={270}, pages={359--371}, year={2007},
 doi={10.1007/s00220-006-0150-x}
}

@article{LowChuang2019,
 author={Low, Guang Hao and Chuang, Isaac L.}, title={Hamiltonian Simulation by Qubitization},
 journal={Quantum}, volume={3}, pages={163}, year={2019},
 doi={10.22331/q-2019-07-12-163}
}

@article{Childs2021,
 author={Childs, Andrew M. and Su, Yuan and Tran, Minh C. and Wiebe, Nathan and Zhu, Shuchen},
 title={Theory of {Trotter} Error with Commutator Scaling},
 journal={Phys. Rev. X}, volume={11}, pages={011020}, year={2021},
 doi={10.1103/PhysRevX.11.011020}
}

@article{LiebRobinson1972,
 author={Lieb, Elliott H. and Robinson, Derek W.},
 title={The Finite Group Velocity of Quantum Spin Systems},
 journal={Commun. Math. Phys.}, volume={28}, pages={251--257}, year={1972},
 doi={10.1007/BF01645779}
}

@article{BravyiHastings2006,
 author={Bravyi, Sergey and Hastings, Matthew B. and Verstraete, Frank},
 title={{Lieb--Robinson} Bounds and the Generation of Correlations and Topological Quantum Order},
 journal={Phys. Rev. Lett.}, volume={97}, pages={050401}, year={2006},
 doi={10.1103/PhysRevLett.97.050401}
}

@article{Schuld2021,
 author={Schuld, Maria and Sweke, Ryan and Meyer, Johannes Jakob},
 title={Effect of Data Encoding on the Expressive Power of Variational Quantum-Machine-Learning Models},
 journal={Phys. Rev. A}, volume={103}, pages={032430}, year={2021},
 doi={10.1103/PhysRevA.103.032430}
}

@article{HuangEtAlNature2021,
 author={Huang, Hsin-Yuan and Broughton, Michael and Mohseni, Masoud and Babbush, Ryan and Boixo, Sergio and Neven, Hartmut and McClean, Jarrod R.},
 title={Power of Data in Quantum Machine Learning},
 journal={Nat. Commun.}, volume={12}, pages={2631}, year={2021},
 doi={10.1038/s41467-021-22539-9}
}

@article{Valiant1984,
 author={Valiant, Leslie G.}, title={A Theory of the Learnable},
 journal={Communications of the ACM}, volume={27}, number={11}, pages={1134--1142}, year={1984},
 doi={10.1145/1968.1972}
}

@article{Hoeffding1963,
 author={Hoeffding, Wassily},
 title={Probability Inequalities for Sums of Bounded Random Variables},
 journal={J. Am. Stat. Assoc.}, volume={58}, number={301}, pages={13--30}, year={1963},
 doi={10.1080/01621459.1963.10500830}
}

@article{Haussler1992,
 author={Haussler, David},
 title={Decision Theoretic Generalizations of the {PAC} Model for Neural Net and Other Learning Applications},
 journal={Information and Computation}, volume={100}, number={1}, pages={78--150}, year={1992},
 doi={10.1016/0890-5401(92)90010-D}
}

@article{BartlettMendelson2002,
 author={Bartlett, Peter L. and Mendelson, Shahar},
 title={Rademacher and Gaussian Complexities: Risk Bounds and Structural Results},
 journal={J. Mach. Learn. Res.}, volume={3}, pages={463--482}, year={2002},
 url={https://www.jmlr.org/papers/v3/bartlett02a.html}
}

@article{CuckerSmale2002,
 author={Cucker, Felipe and Smale, Steve}, title={On the Mathematical Foundations of Learning},
 journal={Bull. Am. Math. Soc.}, volume={39}, number={1}, pages={1--49}, year={2002},
 doi={10.1090/S0273-0979-01-00923-5}
}

@incollection{Bousquet2004,
 author={Bousquet, Olivier and Boucheron, St{\'e}phane and Lugosi, G{\'a}bor},
 title={Introduction to Statistical Learning Theory},
 booktitle={Advanced Lectures on Machine Learning},
 editor={Bousquet, Olivier and von Luxburg, Ulrike and R{\"a}tsch, Gunnar},
 series={Lecture Notes in Computer Science}, volume={3176}, pages={169--207},
 publisher={Springer}, year={2004},
 doi={10.1007/978-3-540-28650-9_8}
}

@article{MarvianSpekkens2014,
 author={Marvian, Iman and Spekkens, Robert W.},
 title={Modes of Asymmetry: The Application of Harmonic Analysis to Symmetric Quantum Dynamics and Quantum Reference Frames},
 journal={Phys. Rev. A}, volume={90}, pages={062110}, year={2014},
 doi={10.1103/PhysRevA.90.062110}
}

@article{MolteniGyurikDunjko2026,
 author={Molteni, Riccardo and Gyurik, Casper and Dunjko, Vedran},
 title={Exponential Quantum Advantages in Learning Quantum Observables from Classical Data},
 journal={npj Quantum Inf.}, volume={12}, pages={19}, year={2026},
 doi={10.1038/s41534-025-01162-2}
}

@article{CahaLandauNagaj2018,
  author={Caha, Libor and Landau, Zeph and Nagaj, Daniel},
  title={Clocks in Feynman's Computer and Kitaev's Local Hamiltonian: Bias, Gaps, Idling, and Pulse Tuning},
  journal={Phys. Rev. A}, volume={97}, pages={062306}, year={2018},
  doi={10.1103/PhysRevA.97.062306}
}

@misc{Bandyopadhyay2026,
 author={Bandyopadhyay, Rahul and Molteni, Riccardo and Eisert, Jens and Dunjko, Vedran and Jerbi, Sofiene},
 title={Provable Learning Separation for Predicting Time-Evolution of Quantum Many-Body Systems},
 year={2026}, eprint={2607.06472}, archivePrefix={arXiv},
 primaryClass={quant-ph},
 doi={10.48550/arXiv.2607.06472},
 url={https://arxiv.org/abs/2607.06472}
}

@article{ChuLiCai2023,
  author={Chu, Yaoming and Li, Xiangbei and Cai, Jianming},
  title={Strong Quantum Metrological Limit from Many-Body Physics},
  journal={Phys. Rev. Lett.},
  volume={130},
  pages={170801},
  year={2023},
  doi={10.1103/PhysRevLett.130.170801},
  eprint={2301.12113},
  archivePrefix={arXiv},
  primaryClass={quant-ph}
}

@misc{MolteniMarshallDunjko2026,
  author={Molteni, Riccardo and Marshall, Simon C. and Dunjko, Vedran},
  title={Quantum Machine Learning Advantages Beyond Hardness of Evaluation},
  year={2025},
  eprint={2504.15964},
  archivePrefix={arXiv},
  primaryClass={quant-ph},
  note={Revised February 2026, version 2},
  url={https://arxiv.org/abs/2504.15964}
}

@article{JanzingWocjan2005,
  author={Janzing, Dominik and Wocjan, Pawel},
  title={{Ergodic Quantum Computing}},
  journal={Quantum Inf. Process.},
  volume={4},
  pages={129--158},
  year={2005},
  doi={10.1007/s11128-005-4482-9}
}

@article{Baez2020,
  author={Baez, Maria Laura and Goihl, Marcel and Haferkamp, Jonas and Bermejo-Vega, Juani and Gluza, Marek and Eisert, Jens},
  title={{Dynamical structure factors of dynamical quantum simulators}},
  journal={Proc. Natl. Acad. Sci. U.S.A.},
  volume={117},
  pages={26123--26134},
  year={2020},
  doi={10.1073/pnas.2006103117}
}

@article{WildAlhambra2023,
  author={Wild, Dominik S. and Alhambra, {\'A}lvaro M.},
  title={{Classical Simulation of Short-Time Quantum Dynamics}},
  journal={PRX Quantum},
  volume={4},
  pages={020340},
  year={2023},
  doi={10.1103/prxquantum.4.020340}
}

@article{BurgarthMaruyamaNori2009,
  author={Burgarth, Daniel and Maruyama, Koji and Nori, Franco},
  title={{Coupling strength estimation for spin chains despite restricted access}},
  journal={Phys. Rev. A},
  volume={79},
  pages={020305},
  year={2009},
  doi={10.1103/physreva.79.020305}
}

@article{BurgarthMaruyama2009,
  author={Burgarth, Daniel and Maruyama, Koji},
  title={{Indirect Hamiltonian identification through a small gateway}},
  journal={New J. Phys.},
  volume={11},
  pages={103019},
  year={2009},
  doi={10.1088/1367-2630/11/10/103019}
}

@article{ZhangSarovar2014,
  author={Zhang, Jun and Sarovar, Mohan},
  title={{Quantum Hamiltonian Identification from Measurement Time Traces}},
  journal={Phys. Rev. Lett.},
  volume={113},
  pages={080401},
  year={2014},
  doi={10.1103/physrevlett.113.080401}
}

@article{GhoshGilchristBurgarth2026,
  author={Ghosh, Riddhi and Gilchrist, Alexei and Burgarth, Daniel},
  title={{Characterizing errors in parameter estimation by local measurements}},
  journal={Phys. Rev. Research},
  volume={8},
  pages={023120},
  year={2026},
  doi={10.1103/4czy-q998}
}

@article{KwonLieJiang2026,
  author={Kwon, Hyukgun and Lie, Seok Hyung and Jiang, Liang},
  title={{Universal Sample Complexity Bounds in Quantum Learning Theory via Fisher Information Matrix}},
  journal={PRX Quantum},
  volume={7},
  pages={033063},
  year={2026},
  doi={10.1103/8k5v-ddtw}
}

@article{BartlettRudolphSpekkens2007,
  author={Bartlett, Stephen D. and Rudolph, Terry and Spekkens, Robert W.},
  title={{Reference frames, superselection rules, and quantum information}},
  journal={Rev. Mod. Phys.},
  volume={79},
  pages={555--609},
  year={2007},
  doi={10.1103/RevModPhys.79.555}
}

@article{SchuldGradients2019,
  author={Schuld, Maria and Bergholm, Ville and Gogolin, Christian and Izaac, Josh and Killoran, Nathan},
  title={{Evaluating analytic gradients on quantum hardware}},
  journal={Phys. Rev. A},
  volume={99},
  pages={032331},
  year={2019},
  doi={10.1103/PhysRevA.99.032331}
}

@article{Wierichs2022,
  author={Wierichs, David and Izaac, Josh and Wang, Cody and Lin, Cedric Yen-Yu},
  title={{General parameter-shift rules for quantum gradients}},
  journal={Quantum},
  volume={6},
  pages={677},
  year={2022},
  doi={10.22331/q-2022-03-30-677}
}

@article{BaireyAradLindner2019,
  author={Bairey, Eyal and Arad, Itai and Lindner, Netanel H.},
  title={{Learning a Local Hamiltonian from Local Measurements}},
  journal={Phys. Rev. Lett.},
  volume={122},
  pages={020504},
  year={2019},
  doi={10.1103/physrevlett.122.020504}
}

@article{HuangTongFangSu2023,
  author={Huang, Hsin-Yuan and Tong, Yu and Fang, Di and Su, Yuan},
  title={{Learning Many-Body Hamiltonians with Heisenberg-Limited Scaling}},
  journal={Phys. Rev. Lett.},
  volume={130},
  pages={200403},
  year={2023},
  doi={10.1103/PhysRevLett.130.200403}
}

@article{Kubo1957,
  author={Kubo, Ryogo},
  title={{Statistical-Mechanical Theory of Irreversible Processes. I. General Theory and Simple Applications to Magnetic and Conduction Problems}},
  journal={J. Phys. Soc. Jpn.},
  volume={12},
  pages={570--586},
  year={1957},
  doi={10.1143/jpsj.12.570}
}

@article{DawsonNielsen2006,
  author={Dawson, Christopher M. and Nielsen, Michael A.},
  title={{The Solovay-Kitaev algorithm}},
  journal={Quantum Inf. Comput.},
  volume={6},
  pages={81--95},
  year={2006},
  doi={10.26421/qic6.1-6}
}

@article{Cheneau2012LightCone,
  author  = {Cheneau, Marc and Barmettler, Peter and Poletti, Dario and
             Endres, Manuel and Schau{\ss}, Peter and Fukuhara, Takeshi and
             Gross, Christian and Bloch, Immanuel and Kollath, Corinna and Kuhr, Stefan},
  title   = {Light-cone-like spreading of correlations in a quantum many-body system},
  journal = {Nature},
  volume  = {481},
  pages   = {484--487},
  year    = {2012},
  doi     = {10.1038/nature10748},
  url     = {https://doi.org/10.1038/nature10748},
  eprint  = {1111.0776},
  archivePrefix = {arXiv},
  primaryClass  = {cond-mat.quant-gas}
}

@article{Jurcevic2014Quasiparticle,
  author  = {Jurcevic, P. and Lanyon, B. P. and Hauke, P. and Hempel, C. and
             Zoller, P. and Blatt, R. and Roos, C. F.},
  title   = {Quasiparticle engineering and entanglement propagation in a
             quantum many-body system},
  journal = {Nature},
  volume  = {511},
  pages   = {202--205},
  year    = {2014},
  doi     = {10.1038/nature13461},
  url     = {https://doi.org/10.1038/nature13461},
  eprint  = {1401.5387},
  archivePrefix = {arXiv},
  primaryClass  = {quant-ph}
}

@article{Braumuller2022InformationPropagation,
  author  = {Braum{\"u}ller, Jochen and Karamlou, Amir H. and Yanay, Yariv and
             Kannan, Bharath and Kim, David and Kjaergaard, Morten and
             Melville, Alexander and Niedzielski, Bethany M. and Sung, Youngkyu and
             Veps{\"a}l{\"a}inen, Antti and Winik, Roni and Yoder, Jonilyn L. and
             Orlando, Terry P. and Gustavsson, Simon and Tahan, Charles and Oliver, William D.},
  title   = {Probing quantum information propagation with out-of-time-ordered correlators},
  journal = {Nature Physics},
  volume  = {18},
  pages   = {172--178},
  year    = {2022},
  doi     = {10.1038/s41567-021-01430-w},
  url     = {https://doi.org/10.1038/s41567-021-01430-w},
  eprint  = {2102.11751},
  archivePrefix = {arXiv},
  primaryClass  = {quant-ph}
}

@article{Kokcu2024LinearResponse,
  author  = {K{\"o}kc{\"u}, Efekan and Labib, Heba A. and Freericks, J. K. and Kemper, A. F.},
  title   = {A linear response framework for quantum simulation of bosonic and
             fermionic correlation functions},
  journal = {Nature Communications},
  volume  = {15},
  pages   = {3881},
  year    = {2024},
  doi     = {10.1038/s41467-024-47729-z},
  url     = {https://doi.org/10.1038/s41467-024-47729-z},
  eprint  = {2302.10219},
  archivePrefix = {arXiv},
  primaryClass  = {quant-ph}
}

@article{Piccinelli2026CircuitDifferentiation,
  author  = {Piccinelli, Samuele and Tacchino, Francesco and Tavernelli, Ivano and Carleo, Giuseppe},
  title   = {A circuit-differentiation framework for {Green}'s functions on quantum computers},
  journal = {Quantum},
  volume  = {10},
  pages   = {2060},
  year    = {2026},
  doi     = {10.22331/q-2026-04-13-2060},
  url     = {https://doi.org/10.22331/q-2026-04-13-2060},
  eprint  = {2505.05563},
  archivePrefix = {arXiv},
  primaryClass  = {quant-ph}
}

@article{JordanWigner1928,
  author = {Jordan, Pascual and Wigner, Eugene},
  title = {{{\"U}ber das Paulische {\"A}quivalenzverbot}},
  journal = {Zeitschrift f{\"u}r Physik},
  volume = {47},
  number = {9--10},
  pages = {631--651},
  year = {1928},
  doi = {10.1007/BF01331938}
}

@incollection{Klich2003,
  author = {Klich, Israel},
  title = {An Elementary Derivation of {Levitov}'s Formula},
  booktitle = {Quantum Noise in Mesoscopic Physics},
  editor = {Nazarov, Yuli V.},
  series = {NATO Science Series II: Mathematics, Physics and Chemistry},
  volume = {97},
  pages = {397--402},
  publisher = {Kluwer Academic Publishers},
  address = {Dordrecht},
  year = {2003},
  doi = {10.1007/978-94-010-0089-5_19},
  eprint = {cond-mat/0209642},
  archivePrefix = {arXiv}
}

@article{Robbins1955,
  author = {Robbins, Herbert},
  title = {A Remark on {Stirling}'s Formula},
  journal = {The American Mathematical Monthly},
  volume = {62},
  number = {1},
  pages = {26--29},
  year = {1955},
  doi = {10.2307/2308012}
}

@book{BoydVandenberghe2004,
  author = {Boyd, Stephen and Vandenberghe, Lieven},
  title = {Convex Optimization},
  publisher = {Cambridge University Press},
  address = {Cambridge},
  year = {2004},
  doi = {10.1017/CBO9780511804441}
}

@article{vonNeumann1928,
  author = {von Neumann, John},
  title = {{Zur Theorie der Gesellschaftsspiele}},
  journal = {Mathematische Annalen},
  volume = {100},
  pages = {295--320},
  year = {1928},
  doi = {10.1007/BF01448847}
}

@inproceedings{Adleman1978,
  author = {Adleman, Leonard},
  title = {Two Theorems on Random Polynomial Time},
  booktitle = {19th Annual Symposium on Foundations of Computer Science},
  pages = {75--83},
  publisher = {IEEE},
  year = {1978},
  doi = {10.1109/SFCS.1978.37}
}

@article{AlMohyHigham2011,
  author = {Al-Mohy, Awad H. and Higham, Nicholas J.},
  title = {Computing the Action of the Matrix Exponential, with an Application to Exponential Integrators},
  journal = {SIAM Journal on Scientific Computing},
  volume = {33},
  number = {2},
  pages = {488--511},
  year = {2011},
  doi = {10.1137/100788860}
}

@dataset{lyu_2026_23668910,
  author       = {Lyu, Yuting},
  title        = {Data of BQP-Complete Susceptibilities Observable
                   at Ballistic Times in Local Quantum Dynamics
                  },
  month        = oct,
  year         = 2026,
  publisher    = {Zenodo},
  doi          = {10.5281/zenodo.23668910},
  url          = {https://doi.org/10.5281/zenodo.23668910},
}
\end{document}